# Exploration of strongly correlated states in SmB$_6$ through a comparison of its two-coil pick-up response to that of Bi$_2$Se$_3$

Sayantan Ghosh[1], Sugata Paul[1], Amit Jash[1,2], Zachary Fisk[3], S. S. Banerjee[1,†]

[1]*Indian Institute of Technology Kanpur, Kanpur, Uttar Pradesh 208016, India.*

[2]*Department of Condensed Matter Physics, Weizmann Institute of Science, Rehovot, Israel.*

[3]*Department of Physics and Astronomy, University of California at Irvine, Irvine, CA 92697, USA.*

Corresponding author email : [†]satyajit@iitk.ac.in

**Abstract**

Earlier studies on the Kondo insulator SmB$_6$ reveal the presence of a bulk Kondo insulating gap between 30 - 50 K, and the emergence of a conducting surface state only below 4 K. Here, we compare the two-coil mutual inductance pick-up response of SmB$_6$ single crystal with that of a conventional topological insulator (TI), Bi$_2$Se$_3$ single crystal. From these studies we identify three distinct temperature regimes for SmB$_6$, viz., (i) $T \geq T^*$(~ 66 K), (ii) (40 K~) $T_g \leq T < T^*$, and (iii) $T < T_g$. At $T^*$ in SmB$_6$, we observe a peak in the temperature-dependent AC pickup signal which corresponds to the peak in the broad hump feature in the bulk DC susceptibility measurements and features in the resistivity measurements. A dip in the pickup signal at $T_g$ in SmB$_6$ correlates with the evidence for the opening of a bulk Kondo gap in transport measurements. Our study of the pickup signal in SmB$_6$ suggests the presence of a thin (submicron order thickness) high conducting surface layer from a temperature just below $T_g$. In this $T$ regime in SmB$_6$, the pickup signal shows a distinct square root frequency ($f$) dependence compared to the linear $f$ dependence found in Bi$_2$Se$_3$. Across all the different $T$ regimes, distinct AC frequency dependence and scaling properties are observed. Our results suggest that above $T^*$, weak exchange interactions cause electrons to scatter from random ion sites. Electronic correlations gradually strengthen with the onset of Kondo like hybridization, setting in from below $T^*$, and at $T_g$, a strongly correlated Kondo gap opens up in the bulk of the material. The appearance of the thin high conducting surface layer is nearly coincident with the onset of bulk Kondo insulating state below $T_g$ in SmB$_6$.

**INTRODUCTION**

Samarium Hexaboride ($SmB_6$) belongs to the Kondo lattice insulator family where the itinerant $d$ – states of Sm ions hybridize with its localized $f$ – states to create a quenched lattice of screened $f$ – electron moments. This leads to a correlated Kondo insulating state with a bulk Kondo gap [1-3], corresponding to a temperature scale $T_g$. Most reports suggest that $T_g$ ranges between 30 K to 50 K [4-6] in $SmB_6$. Resistivity ($\rho$) vs. temperature ($T$) [4-6] behavior in $SmB_6$ shows a significant increase in $\rho$ with lowering of temperature $(T) < T_g$. However, the rapid rise in $\rho$ with decreasing $T$ appears to saturate below 4 K. This observation suggested the presence of low temperature ($T$) conduction channel in $SmB_6$, which has triggered numerous debates [7]. Measurements reveal that the saturation feature in $\rho$ is independent of sample quality [4-6]. Unlike a typical band insulator, $SmB_6$ possesses strongly interacting electronic states along with a significant spin-orbit interaction. Theoretical calculations suggest that these spin-orbit interactions due to hybridization between the conduction band and $f$ – electrons result in gapless surface excitations with time-reversal symmetry [8,9]. Such studies imply that $SmB_6$ belongs to a distinct category of strongly interaction-driven topological material [10-13]. However, the scenario emerging from experiments is complex. Transport studies [14-17] see a saturation of resistivity at low $T$. These studies together with photoemission spectroscopy [18-21] and neutron scattering measurements [22] suggest the saturation could be due to the presence of a conducting surface state in $SmB_6$ at low temperatures i.e. below 4 K, with a bulk insulating gap as seen from point contact spectroscopy [23]. While some measurements [24] suggest that at low $T$, de Haas–van Alphen (dHvA) oscillations correspond to conduction from 2D surface states; some other measurements, however, claim the oscillations to originate from bulk [25,26]. The dHvA quantum oscillation measurements [24,26,27] and ARPES [20,28,29] show that the measured effective mass of electrons at low $T$ is low, whereas, STM measurements suggest a high effective mass [30]. A common feature of these studies is that, in $SmB_6$ there is the opening of a Kondo gap in the bulk at $T_g \sim$ 30 K to 50 K, with evidences suggesting the emergence of a conducting surface state only at a much lower $T$ (< 4 K). Thus $SmB_6$ is an exotic Kondo insulator (KI) with prospects of having topological surface states at low $T$ [7,31-35]. In the present paper, by using the sensitive two-coil mutual inductance technique for $SmB_6$, we explore the $T$ regime around $T_g$ and attempt to identify features of conductivity emerging in the background of a bulk Kondo gap below $T_g$.

While electrical transport studies are used to explore TI materials [13,19,36-41], however in these measurements, the parallel surface and bulk conducting channel contributions to conductivity are frequently admixed. In the non-contact type two-coil mutual inductance technique [42,43], it was shown that in conventional TI material $Bi_2Se_3$, the bulk and topological surface states give rise to unique and distinct frequency ($f$) and $T$ dependent features. One sees a linear frequency dependence in the temperature regime where topological surface states contribute to the electrical conductivity while a quadratic $f$ dependence is observed where bulk conductivity dominates [42,43]. In this paper, using the above technique, we compare the behavior of single crystal of $SmB_6$ with a well known non-interacting TI material, viz., a single crystal of $Bi_2Se_3$ [44,45]. We measure the in-phase pick-up signal ($V'$) for the two single crystals at different temperature ($T$) and excitation frequency ($f$). $Bi_2Se_3$ at low $f$, below 100 K shows thermally activated bulk conductivity. For $SmB_6$, we identify another temperature scale $T^*$ above $T_g$. We see that here the $T$ dependent pick-up signal $V'(T)$ can be separated into three distinct regimes: regime (i) : $T \geq T^* \sim$ 66 K; regime (ii) : (~ 40 K) $T_g \leq T < T^*$; regime (iii) : $T < T_g$. Around $T^*$ in $SmB_6$, we observe a peak in the temperature-dependent AC pick-up signal which corresponds to the peak in the broad hump feature in the bulk DC susceptibility measurements. Near $T^*$ we also see a non-linear dependence of $V'$ signal on $f$ and the excitation field amplitude as well. Such features are absent in the pick-up signal at $T < T_g$ and at $T > T^*$. Bulk electrical conductivity measurement in our $SmB_6$ single crystal shows a featureless slow increase in resistivity (decrease in conductivity) in regime (i). However, in regime (ii) the rate of increase in resistivity becomes slightly significant and at regime (iii) i.e., below $T_g$ there is a rapid increase in resistivity which is associated with the opening of a Kondo insulating gap in the bulk. A detailed comparison of $V'(T)$ response at low $T$ in $SmB_6$ and $Bi_2Se_3$, suggests the appearance of a thin high conducting surface layer, from the vicinity of $T_g$ in $SmB_6$, giving rise to an excess conduction while the bulk remains insulating. In this regime for $SmB_6$, we find the frequency dependent pick-up response $V'(f) \propto f^{\frac{1}{2}}$ compared to a linear $f$ dependence found for the conducting surface state feature in $Bi_2Se_3$. In the bulk-dominated conductivity regime, viz., well above $T_g$ in $SmB_6$ and above 70 K in $Bi_2Se_3$, both materials exhibit identical $f^2$ dependence of $V'$. The $f^{\frac{1}{2}}$ dependence is probably a result of the presence of strong $f-f$ electron correlations modifying the features of the conducting surface layer in $SmB_6$. Results of our investigations suggest that above $T^*$, weak exchange interactions cause electrons to scatter from random ion sites. Electronic correlations gradually strengthen below $T^*$ as Kondo hybridization begins and finally, a strongly correlated bulk gap appears at $T_g$. Our results suggest, the appearance of the thin high conducting surface layer is nearly coincident with the onset of bulk Kondo insulating state below $T_g$ in $SmB_6$.

**RESULTS**

**Two-coil measurement technique**

The $SmB_6$ single crystal grown by the Aluminium flux method [25,35] has dimensions of 0.9 mm × 0.75 mm × 0.2 mm. The $Bi_2Se_3$ single crystal prepared by slow cooling stoichiometric melts of high purity Bismuth (Bi) and Selenium (Se) powders [40,41,46] has dimensions of 3.9 mm × 2.5 mm × 0.069 mm. Figure 1(a) shows a schematic of our non-contact type two-coil mutual inductance measurement setup [42,43] where $SmB_6$ is kept between the excitation and pick-up coil. The physical dimensions, the number of turns and inductance of the two coils are closely matched (see section I in supplementary). Alternating current ($I$) at frequency $f$ sent in the excitation coil, creates a time-varying magnetic field. This AC excitation magnetic field is experienced by the sample placed between the two coils. The magnetic response generated by the sample, in turn, induces a voltage in the pick-up coil, which is measured using a Stanford SR830 DSP lock-in amplifier. We have used a Janis closed cycle cryostat for low temperature measurements. A 1.5 mm thick oxygen-free high conductivity (OFHC) Copper sheet (with an insulating coat) with a hole of diameter 0.75 mm is placed on top of the excitation coil and the sample is placed just above the hole to reduce the background pick-up signal [42] generated by the stray alternating magnetic field outside the sample (see Fig. 1(a)). The OFHC - Cu sheet with a conductivity of ~ $6 \times 10^7$ S.m$^{-1}$, has a skin depth ($\delta$) of ~ 200 μm for a 1 kHz AC field. Because the thickness of our OFHC - Cu sheet is greater than $\delta$, most of the AC field outside the sample is shielded out except over the hole [42]. The hole concentrates the AC magnetic field directly onto the sample placed just over the hole [42], and thus the pick-up signal is predominantly contributed by the sample. This has been confirmed by finite element analysis using COMSOL Multiphysics software simulations reported in our earlier work [42]. The stray AC fields, however, constitute a significant fraction of the background signal in the pick-up coil. Note that for every measurement, we subtract the background signal from total response such that the pick-up signals contain solely the sample response. Only the Cu sheet with a hole, placed between the coils is used to measure the background. If $I_0$ is the peak amplitude of the AC excitation current in the excitation coil, then the induced pick-up voltage is, $V(f) = I_0 M f$, where $M$ is the mutual inductance that contains the sample response. Note that $M$ is proportional to the AC susceptibility ($\chi_{ac}$) of the sample as $V = M\left(\frac{di_{ac}}{dt}\right)$ and $V = -Nk\chi_{ac}\left(\frac{dh_{ac}}{dt}\right)$ where $N$ is the number of turns of the pick-up coil, $k$ is the geometric filling factor and $\chi_{ac} = \chi' + i\chi''$. The AC pick-up voltage $V = V' + iV''$, where in this paper we will discuss the behavior of the in-phase signal, $V'$. It is also known that for a conducting sample, the AC field from the excitation coil induces currents in the sample whose strength is proportional to the $f$ dependent electrical conductivity ($\sigma$) of the material. These currents in the sample induce a signal in the pick-up coil. Hence as shown through Eq. 1 in ref. [42] the pick-up signal is also related to $\sigma(f)$ of the material. Therefore, the pick-up signal in the two-coil technique not only depends on $\chi$ but also on $\sigma$. While Bi$_2$Se$_3$ is non-magnetic, there is a discernible $\chi$ response for SmB$_6$. Using the SQUID magnetometer, we measure the $\chi(T)$ for SmB$_6$ over the $T$ regime of interest. Normalizing the $V'(T)$ response with $\chi(T)$, we obtain a signal which is representative of the behavior of $\sigma(T)$ of the sample (details discussed subsequently). Another benefit of measuring the conductivity of the sample using the two-coil technique is that, depending on the $f$ of the excitation AC field, signals acquired closer to the conducting sample surface can be distinguished from those obtained from deeper within the sample volume. As low $f$ excitation signal penetrates deep into a conducting sample (since $\delta \propto \frac{1}{\sqrt{f\sigma}}$), it leads to a pick-up signal representing information from a larger sample volume (which we will subsequently refer to as the bulk sample response) compared to higher $f$, where the field penetrates into a thin layer (submicron order thickness) close to the sample surface (see estimate of skin depth ($\delta$) for Bi$_2$Se$_3$ in section V of Ref. [42], and for SmB$_6$ discussed later here). This feature was previously used to distinguish between the behavior of bulk ($\sigma_b$) and surface ($\sigma_s$) electrical conductivity in a non-interacting TI, Bi$_2$Se$_3$ single crystal [42,43,47]. We use the above technique now to compare the $V'$ behavior of single crystals of SmB$_6$ and Bi$_2$Se$_3$.

**Comparison of SmB$_6$ and Bi$_2$Se$_3$ $V'(T)$ response**

Figure 1(b) shows $V'(T)$ for SmB$_6$ and Bi$_2$Se$_3$ single crystals measured at $f = 4$ kHz excitation frequency, with $I_0 = 200$ mA applied to the excitation coil. For Bi$_2$Se$_3$ (blue symbols), the $V'(T)$ decreases below 100 K. Fig 1(b) inset depicts a linear (see red dashed line) $\ln V'(T)$ vs. $T^{-1}$ behavior, which is a feature of activated charge conductivity in bulk of Bi$_2$Se$_3$. At 4 kHz, the $\sigma_b$ in Bi$_2$Se$_3$, has a form, $V'(T) \propto \sigma_b(T) \propto \exp\left(-\frac{\Delta}{k_B T}\right)$, (where $\Delta$ is the activation energy scale). The activated behavior of bulk conductivity in Bi$_2$Se$_3$ is produced due to the doped charges created by Se vacancies in the crystal [48]. We estimate the activation energy scale $\Delta$ is ~ 5.92 ± 0.31 meV from the slope of the red dashed line in Fig. 1(b) inset, which is consistent with previous results [42]. It is clear that SmB$_6$ crystal's $V'(T)$ is completely different. In Fig. 1(b), we see that with reducing $T$ from 100 K, the $V'(T)$ of SmB$_6$ initially increases and exhibits a peak at $T^* \sim$ 66 K. Below $T^*$, the $V'(T)$ rapidly decreases to reach a minima at $T_g \sim$ 40 K (see Fig. 1(b)). Below $T_g$, the $V'(T)$ increases once again to finally saturate at low $T$ for SmB$_6$.

**Transport and DC magnetization measurements of SmB$_6$**

Figure 2(a) shows our standard four-probe measurements of resistivity ($\rho$) versus $T$ for the SmB$_6$ single crystal. We observe a monotonic increase in $\rho$ with decreasing $T$. Approximately the $\rho(T)$ shows a weak $T$ dependent regime from high $T$ down to ~ 65 K. Below 65 K the $\rho(T)$ increases significantly and shows exponential rise from $T_g \sim$ 40 K. This behavior of $\rho(T)$ is consistent with earlier studies in SmB$_6$ [4,5,14,15,49,50], and the value of $T_g$ is also in a similar range [1,15,30]. Inset (I) of Fig. 2(a) shows that $\rho(T)$ obeys the conventional Hamann fit [51] down to $T^* \sim$ 65 K and then starts to deviate from it. This has been elaborated upon in the discussion section later. Note that the minima in $V'(T)$ of SmB$_6$ (see Fig. 1(b)) occurs at the Kondo gap temperature $T_g$. Inset II of Fig. 2(a) shows the comparison of the effective conductivity derived from normalised pick-up response $V'(T)/\chi(T)$ and the bulk conductivity ($\sigma(T)$) response from transport (see the comparison up to 300 K in supplementary section VI). We observe that from 80 K to 40 K (= $T_g$), while the electrical conductivity ($\sigma$) changes by small amount, the $V'(T)/\chi(T)$ drops significantly. At $T < T_g$, while the bulk $\sigma$ drops rapidly due to formation of the Kondo gapped state, the $V'(T)/\chi(T)$ shows a significant increase. Note that the size of the cusp like feature at $T^*$ is significantly diminished compared to the rise in $V'(T)/\chi(T)$ seen below $T_g$. This difference suggests, the enhancement in effective conductivity is from the thin surface layer below $T_g$ in SmB$_6$. Although the feature in $V'(T)$ at $T^*$ (in Fig. 1(b)) is not prominent in $V'(T)/\chi(T)$ (Fig. 2(a) inset II), it appears as a modulation in $\rho(T)$ behavior, which is seen more clearly from the behavior of the absolute value of $\frac{d\rho}{dT}$ versus $T$ (log-log scale) around $T^*$ = 65 K ± 2 K (see section III in supplementary). A similar change is also seen near $T_g$ = 40 K ± 2 K (see section III in supplementary). At $T_g$, the strongly correlated Kondo insulating gap opens in the bulk where the resistivity increases sharply at $T < T_g$.

The 5 Tesla isofield DC $\chi(T)$ measurement (cryofree Cryogenics UK, SQUID magnetometer) for SmB$_6$ single crystal is shown in Fig. 2(b) inset (blue data points). Below 300 K there is a monotonic increase in $\chi(T)$ as $T$ is reduced and it has a broad dome shaped characteristic below 100 K. We show that $\chi(T)$ behavior for 5 Tesla zero field cooled (ZFC) and field cooled (FC) conditions are identical, suggesting the absence of any irreversible magnetic component in SmB$_6$ (see section IV in supplementary). Below 15 K, an increase in $\chi(T)$ is observed. Similar nature of $\chi(T)$ as in Fig. 2(b) inset was also seen earlier [52]. The broad dome feature in $\chi(T)$ is usually considered to be related to the opening of a Kondo gap in the bulk of SmB$_6$ [35,52]. However, the increase of $\chi(T)$ below 15 K has been suggested to be related to some fluctuating moments in bulk of SmB$_6$ [52] or the presence of paramagnetic impurities [53]. Note our $T$ regime of interest in the present paper is above 15 K. From the $\frac{d\chi}{dT}$ versus $T$ plot in Fig. 2(b) inset (red data points), we see extrema $\left(\frac{d\chi}{dT} = 0\right)$ at $T^*$ ~ 60 K ± 1 K and at 15.5 K (shown by black arrows). The decrease in $\chi(T)$ below $T^*$ (Fig. 2(b) inset) coincides with a decrease in the pickup signal $V'(T)$ at 4 kHz below $T^*$ (Fig. 1(b)). We do not see any distinguishing feature related to $T_g$ either in $\chi(T)$ or in $\frac{d\chi}{dT}$. The table I below shows that the $T^*$ and $T_g$ obtained from the two-coil measurements (Fig. 1(b)), transport ($\rho(T)$ in Fig. 2(a)) and DC susceptibility ($\chi(T)$ in Fig. 2(b) inset) measurements are all in the similar range with variations of 3 K to 5 K. The variations are attributed to differences in cryogenic conditions in the three different setups used for our measurements. The inverse susceptibility $\chi^{-1}$ vs $T$ behavior for both 5 T ZFC and FC conditions are plotted in Fig. 2(b) main panel. We see that while at high $T$ (above 100 K) the data fits (blue dashed line) to Curie Weiss Law, with $\chi(T) = \frac{C}{T-\theta}$, with $\theta$ = -267 K and Curie constant $C$ = 0.039 emu.K/g.Oe, the data deviates significantly at low $T < 100$ K. Similar nature for the high temperature $\chi(T)$ for SmB$_6$ has also been reported earlier [50]. The negative value of $\theta$ signifies absence of magnetic ordering in SmB$_6$. Note that in other systems, similar Curie Weiss Law is observed in strongly correlated itinerant electron scenario without any long-range magnetic ordering [50,54-56]. In Fig. 2(b) we see that with lowering of $T$ there is a strong deviation from the Curie Weiss nature. A linear back extrapolation (dashed violet line) from the low $T$ where $\chi^{-1}$ has deviated significantly from the Curie Weiss Law, intersects with the back extrapolated Curie Weiss fit from high $T$ regime. The intersection gives us a temperature scale ~ 60 K, below which it appears that the correlations start to develop in the system. Note that this $T$ ~ 60 K is close to our $T^*$ value identified earlier. The above Curie Weiss like nature at high $T$ and deviation from it at low $T$, suggests the transition from an uncorrelated paramagnetic response regime to a strong electronic correlation dominated regime in $SmB_6$ below 60 K.

**Table I: $T^*$ and $T_g$ values determined from different techniques**

| Technique | $T^*$ (K) | $T_g$ (K) |
| --- | --- | --- |
| Two-coil | 66 K ± 1 K | 40 K ± 1 K |
| Resistivity | 65 K ± 2 K | 40 K ± 1 K |
| DC susceptibility | 60 K ± 1 K | -- |

**Comparison of surface conductivity of Bi$_2$Se$_3$ and SmB$_6$**

At $f = 4$ kHz, for $SmB_6$ the $V'(T)$ in Fig. 1(b) increases for $T < T_g$. The minima at $T_g$ in $SmB_6$ is similar to the $V'(T)$ behavior in non-magnetic Bi$_2$Se$_3$, albeit in Bi$_2$Se$_3$, this minima was seen in $V'(T)$ data taken at $f = 65$ kHz [42]. In Figure 3(a), we compare the $V'(T)/\chi(T)$ data for SmB$_6$ at 4 kHz with $V'(T)$ data for Bi$_2$Se$_3$ single crystal measured at $f = 65$ kHz. Recall that Bi$_2$Se$_3$ has no inherent magnetism, therefore the $V'(T)$ behavior is due to its electrical conductivity $\sigma(T)$ response. In Bi$_2$Se$_3$, $\sigma(T) = \sigma_b(0) \exp\left(-\frac{\Delta}{k_B T}\right) + \frac{\sigma_s(0)}{[C+DT]}$, where $\sigma_b(0)$ and $\sigma_s(0)$ are the temperature independent bulk and topological surface state conductivities respectively, $D$ is related to electron-lattice scattering and $C$ is related to electron-electron (*e-e*) interaction strength [38,42]. Due to the small value of the skin depth at high $f = 65$ kHz, only a thin layer around the sample surface in the Bi$_2$Se$_3$ is probed by the excitation field. Hence in Fig. 3(a), at 65 kHz below 70 K, the contribution from the $\exp\left(-\frac{\Delta}{k_B T}\right)$ dependence of bulk conductivity falls below the contribution from the $T$ dependent topological surface state conductivity. This results in the upturn feature in $V'(T)$ which serves as a signature of the contribution from TI surface conducting states. Fig. 3(a) shows that in the $T$ range 50 -70 K, the $V'(T)$ of Bi$_2$Se$_3$ fits with $\frac{1}{C+DT}$ behavior as $V'(T) \propto \sigma_s \propto \frac{1}{C+DT}$. Note that a similar $V'(T) \propto \frac{1}{C+DT}$ dependence has been reported in other crystals of Bi$_2$Se$_3$ via pick-up coil measurements [42] and transport measurements [38]. From Fig. 3(a), we see that below 50 K, the behavior of $V'(T)$ for Bi$_2$Se$_3$ fits to a form $[A - BT]$ (see magenta dashed line). In Figs. 3(b) and 3(c) we plot the value of the fit error at different $T$, $\epsilon_{(C+DT)^{-1}}(T)=(V'(T)$ or $V'(T)/\chi(T)$ calculated using $\frac{1}{C+DT}$) $- (V'(T)$ or $V'(T)/\chi(T)$ measured) and $\epsilon_{(A-BT)}(T) =$ $(V'(T)$ or $V'(T)/\chi(T)$ calculated with $A - BT) - (V'(T)$ or $V'(T)/\chi(T)$ measured), respectively. For Bi$_2$Se$_3$, Fig. 3(c) shows that $\epsilon_{(C+DT)^{-1}} \approx 0$ between 70 K down to 50 K, while below 50 K, $\epsilon_{(A-BT)} \approx 0$. This suggests $(C + DT)^{-1}$ is a good fit to $V'(T)$ data from 70 K to 50 K while below 50 K $[A - BT]$ is a better fit to the data of Bi$_2$Se$_3$. It has been suggested [42,57,58] that, enhanced influence of *e-e* interaction effects at low $T$ is a plausible source for the observed changes in $T$ dependence of $V'$ leading to even a low $T$ saturation of $V'(T)$, as observed in Fig. 3(a). Note that in SmB$_6$ for $T < T_g$, the upturn in $V'(T)/\chi(T)$ cannot be ascribed to any bulk conductivity ($\sigma$) features, since the inset (II) of Fig. 2(a) has already shown the rapid decrease in $\sigma$ due to Kondo localization below $T_g$. Furthermore, Fig. 2(b) inset also indicates that below $T^*$ (and also below $T_g \sim 40$ K) the DC $\chi(T)$ decreases. The upturn in $V'(T)$ below 70 K in Bi$_2$Se$_3$ at high $f = 65$ kHz, had suggested the presence of high conducting surface state features. Observation of a similar upturn in $V'(T)/\chi(T)$, below $T_g$ in SmB$_6$, suggests the presence of a thin high conducting surface layer in this $T$ regime. Figure 3(b) shows that for SmB$_6$, $\epsilon_{(C+DT)^{-1}} \approx 0$ over about 2 K below $T_g = 40$ K. Below 38 K, $\epsilon_{(C+DT)^{-1}}$ deviates significantly from 0 and in this regime $\epsilon_{(A-BT)} \approx 0$. From the above, we see that while the $T$ dependence of $V'$ in Bi$_2$Se$_3$ first shows $\frac{1}{[C+DT]}$ behavior at the upturn below ~ 70 K followed by a $[A - BT]$ behavior only below 50 K, however, in SmB$_6$, $V'(T)$ shows a $[A - BT]$ behavior over a much wider $T$ window which begins from just below $T_g$. Presented below is the table II which summarises the $C$, $D$, $A$, $B$ values for SmB$_6$ and Bi$_2$Se$_3$.

**Table II**: **Comparison of $C$, $D$ and $A$, $B$ values between Bi$_2$Se$_3$ and SmB$_6$**

| Sample | C | D | A | B |
| --- | --- | --- | --- | --- |
| Bi$_2$Se$_3$ (65 kHz) | 0.0052 mV$^{-1}$ | 1.86 mV$^{-1}$.K$^{-1}$ | 20.408 mV | 0.249 mV.K$^{-1}$ |
| SmB$_6$ (4 kHz) | 0.0002 mV$^{-1}$ | 0.53 mV$^{-1}$.K$^{-1}$ | 12.739 mV | 0.320 mV.K$^{-1}$ |

**Comparison of $V'(f)$ response of Bi$_2$Se$_3$ and SmB$_6$ with different $I_0$ at different $T$**

We study the frequency dependence of $V'$ to understand the properties of the thin high conducting surface layer in SmB$_6$. Studies on the frequency dependence of $V'$ in TI material like Bi$_2$Se$_3$ [42,43] have already shown that at low $T$, viz., in the $\sigma_s$ dominated regime, $V'(f) \propto f^\alpha$, where $\alpha \sim 1$, while in the high $T$ regime (above 70 K) where $\sigma_b$ dominates, $\alpha \sim 2$. Figure 4(a) shows $V'(f)$ response on log-log scale for both SmB$_6$ and Bi$_2$Se$_3$ at 17 K, measured with a peak amplitude of the excitation current ($I_0$) = 200 mA in the primary excitation coil. Here, the linear nature of $V'(f)$ with slope of 0.976 ± 0.003 for Bi$_2$Se$_3$ suggests $V' \propto f$ (see magenta line). However, for SmB$_6$ at $T$ = 17 K which is below $T_g$, we observe $V' \propto f^{\frac{1}{2}}$ (slope of red line in Fig. 4(a) is 0.464 ± 0.001). The $f$ dependence of $V'(f)$ in this low $T$ regime is strikingly different in the two materials. Inset of Fig. 4(c) shows multiple data sets confirming the $V'(f) \propto f^{\frac{1}{2}}$ behavior at different $T$ of 17 K, 30 K and 35 K, below $T_g$ = 40 K in SmB$_6$ (also see supplementary section IX for more $T$ data). Figure 4(b) compares $V'(f)$ response for $I_0$ = 200 mA of SmB$_6$ and Bi$_2$Se$_3$ at 70 K (viz., at $T > T_g$ for SmB$_6$ and inside the bulk conducting regime for Bi$_2$Se$_3$). Here we see in the log-log plot of $V'(f)$ that Bi$_2$Se$_3$ has a slope: 1.838 ± 0.0742 (dark yellow line) and for SmB$_6$ the slope is: 2.014 ± 0.041 (orange line) for low $f$ i.e. $V' \propto f^2$ observed for both. For higher $f$ at 70 K, we observe a $V'(f) \propto f$ behavior in both Bi$_2$Se$_3$ and SmB$_6$ (magenta lines in Fig. 4(b)). Recall studies in Bi$_2$Se$_3$ showed that linear frequency dependence of $V'(f)$ was seen in the low $T$ regime where $V'(T) \propto [C + DT]^{-1}$, a feature characteristic of $T$ dependence of surface conductivity in non-interacting conventional TI material [38,42]. In SmB$_6$, the $V'(f) \propto f^2$ in the low $f$ regime at 70 K is a characteristic feature of dominant bulk contribution to conductivity as it had been shown earlier in pick-up coil studies of Bi$_2$Se$_3$ [42]. In the high $T$ regime of 70 K ($> T_g$), as bulk contributions to conductivity is admixed with surface conductivity, therefore, at higher $f$, the mean curvature of $V'(f)$ in SmB$_6$ behaves as $V'(f) \propto f$ instead of $f^{\frac{1}{2}}$. Note that, correlations have not started to develop until below $T^*$ in SmB$_6$. The similar surface and bulk admixture at high $T$ was also observed for Bi$_2$Se$_3$ as reported earlier in ref [42]. Only at temperatures below $T_g$, where the bulk contribution to conductivity diminishes significantly, the unique $V'(f) \propto f^{\frac{1}{2}}$ behavior for SmB$_6$ is seen. In the next section, we discuss a possible scenario for this unique $f$ dependence of $V'(f)$ in SmB$_6$. At $T$ = 70 K, we estimate the relative fraction of bulk ($P_b$) and surface ($P_s$) contributions to $V'(f)$ by fitting it to $V'(f) = P_b[V'_b(f)] + P_s[V'_s(f)]$, where $V'_b(f)$ and $V'_s(f)$ are the pick-up response from the bulk and surface respectively. The data fits well only in a limited frequency window where both the surface ($V'_s(f) \propto f$) and bulk ($V'_b(f) \propto f^2$) contributions are present in the behavior of $V'(f)$ (see supplementary section XIII). At 70 K, the fit yields for SmB$_6$, $P_b$ = 70 % and $P_s$ = 30 % whereas for Bi$_2$Se$_3$, $P_b$ = 51 % and $P_s$ = 49 %. It may be noted that using this analysis, the values for $P_b$ and $P_s$, we obtain for Bi$_2$Se$_3$ at 70 K, are consistent with those reported earlier [42].

We mention here a few possible scenarios which may relate to the observed $T$ and $f$ dependence of $V'$ in SmB$_6$ below $T_g$. The total AC conductivity, $\sigma_{total}(\omega = 2\pi f) = \sigma_{DS}(\omega) + \sigma_{e-e}(\omega)$, where, $\sigma_{DS}(\omega) = \frac{\sigma_0}{(1-i\omega\tau)}[1 + \sum_{n=1}^{\infty} \frac{c_n}{(1-i\omega\tau)^n}]$, is the Drude-Smith's (DS) conductivity which is a generalized form of the conventional AC Drude's conductivity ($\sigma_D(\omega)$) behavior incorporating the effects of multiple scattering events [59,60], $\sigma_0$ is the conventional Drude's DC conductivity, $\tau$ is the time interval between two consecutive scattering, $c_n$ represents the fraction of the carrier's original velocity that is retained after the $n^{th}$ scattering ($n$ is an integer) and the $\sigma_{e-e}(\omega)$ is the conductivity in the presence of strong electron-electron interaction effects. The first term in $\sigma_{DS}(\omega)$ is $\sigma_D(\omega)$. Note that higher order contribution from scattering ($n \geq 1$) in $\sigma_{DS}(\omega)$, have progressively decreasing weights and consequently much weaker contribution to the surface conductivity. While the linear $f$ dependence of $V'$ at low $T$ in Bi$_2$Se$_3$ was explained earlier [42] by considering topological surface state contributions (viz., $\omega\tau \gg 1$) to the leading order term in $\sigma_{DS}(\omega)$, however $V' \propto f^{\frac{1}{2}}$ behavior in SmB$_6$ below $T_g$ can not be explained by the higher order terms in $\sigma_{DS}(\omega)$ alone. In this context, one possible scenario to consider is, the effect of strong *e-e* interactions [61-64], which are known to produce a non integral $f$ dependence of $\sigma_{e-e}(\omega = 2\pi f)$ of the type $\sigma_{e-e}(\omega) \propto \left[ln\left(\frac{2I_0}{\hbar\omega}\right)\right]^3 \left(\frac{\omega}{\epsilon}\right)$. At $T < T_g$, where we observe $V' \propto f^{\frac{1}{2}}$ behavior, we also see the unusual $[A - BT]$ behavior in $V'(T)$. It is possible that in SmB$_6$, these features are associated with emergence of a high conducting surface layer [9,14,23,28,65-67] together with strong $f - f$ electron correlations effects. We would like to mention that the above is suggested as a probable scenario. Additional theoretical studies are needed to validate the applicability of this scenario to SmB$_6$ and also explore other mechanisms to explain the $f^{\frac{1}{2}}$ behavior.

We estimate of the typical thickness of the thin high conducting surface layer seen below $T_g$ in SmB$_6$. The skin depth $\delta = \sqrt{\rho/\pi f \mu}$, where $\mu = \mu_0(1 + \chi)$, depends on $f$ as well as $T$ dependent $\rho$ and $\chi$ of the sample. For SmB$_6$ using our measured values of bulk resistivity, $\rho = 10^{-3}$ Ω.cm and $\chi = 6.7 \times 10^{-3}$ at 100 K (supplementary section IV), we estimate $\delta_{SmB_6}$(100 K) ~ 25 mm at $f$ = 4 kHz. Note that $V'(T) \propto \Lambda(T)$, with $\Lambda(T) \sim s.\delta(T)$, where $\Lambda$ is the effective sample volume which is being probed by the penetrating electromagnetic signal and $s$ is the surface area of the sample. Note that $\Lambda$ has the same $T$ dependence as $\delta$. It is clear that at 100 K the $V'$ response is from the entire bulk of the sample. For SmB$_6$ the estimated $\delta$(100 K) ~ 25 mm at 100 K is much greater than the sample thickness, $t$ of 200 μm, therefore at 100 K, $\Lambda$ (100 $K$) = $s.t$. Using $V'(T) \propto \Lambda(T)$, for SmB$_6$, we estimate at $f$ = 4 kHz, $\delta_{SmB_6}(40\ K) = \frac{V'(40\ K)}{V'(100\ K)} \times t$ = 4.7 μm, where $V'(T)$ are the directly measured values. Note the above method for estimating $\delta_{SmB_6}$ works down to the $T$ regime of 40 K where there is a dominant bulk contribution to electrical conductivity, i.e., at $f$ = 4 kHz where we see $V'(f) \propto f^2$ behavior. Below 40 K the onset of higher conductivity surface layer in SmB$_6$ ($[A - BT]$ dependence of $V'(T)$ and $V'(f) \propto f^{\frac{1}{2}}$ behavior) screens the probing signal from most of the bulk. Therefore below 40 K, to estimate $\delta$, we consider surface conductivity $\sigma_s(T) = \sigma_{0s}(A - BT)$, where $\sigma_{0s}$ is a constant $T$ independent proportionality factor and $A, B$ values are determined from the linear fit to the effective conductivity response of SmB$_6$ at 4 kHz (see Fig. 3(a)). Using $\delta = \delta_0 \sqrt{\frac{\sigma_0}{\sigma_s(T)}}$, where $\delta_0 = \delta_{SmB_6}(40\ K)$ and $\sigma_0 = \sigma_{SmB_6}^{40\ K}$, we estimate $\delta$ ~ 400 nm below 25 K (see supplementary section XII for detailed calculation and plot of $T$ dependence of estimated $\delta$ for SmB$_6$ and Bi$_2$Se$_3$). Therefore, in SmB$_6$ for a 4 kHz pickup measurement at low $T$, the contribution to the pickup signal is predominantly from about 400 nm thick surface layer. Hence, the thin high conducting surface layer we are probing at $T < T_g$ is of sub-micron order thickness. Here we would like to mention that we have not polished any of our crystal faces before any of our pick-up coil measurements. Also note that the behavior of $V'(T)$ around $T_g$ doesn't change significantly for measurements done at larger $f$ (see plot of $V'(T)$ at $f$ = 4, 8, 12 and 55 kHz in supplementary section VIII). Electrical transport studies in SmB$_6$ thin films [68,69] suggest, surface conducting features appear only below 10 K. It is known that in electrical transport measurements there is mixing of the surface and bulk conductivities in the net measured conductivity due to parallel transport channels through the bulk and surface. Disentangling and determining the two contributions is fairly complex issue in such measurements. Here only at very low $T$ (4 K and below), as the conducting channels through the bulk become almost open due to high resistance, especially in $SmB_6$, does one begin to observe the surface contribution to electrical conductivity. Due to these complications, it may not always be possible to determine at what $T$ does the onset of surface conductivity occur in electrical transport measurements. However, in our $f$ dependent pickup coil technique we see the presence of a thin high conducting surface layer appear at $T < T_g$ in $SmB_6$.

Recall that the AC field from the excitation coil which the sample experiences is proportional to the amplitude of the AC current ($I_0$) in the excitation coil. We plot $\frac{V'}{I_0}$ vs. $f$ in Figs. 4(c), 4(d) and 4(e) for different $I_0$ and at $T$ values chosen in regimes labelled (iii), (ii) and (i) in Fig. 1(b) respectively. The $V'(f)$ signal scales with $I_0$ in the regime (iii) i.e. $T \ll T_g$ (Fig. 4(c)) and also in the regime (i) i.e. $T > T^*$ (see Fig. 4(e)). Inset of Fig. 4(c) shows the scaled behavior of $V'(f)$ for different $I_0$ at different $T$ below $T_g$. However, in regime (ii), $T_g < T < T^*$ (see Fig. 4(d)) for $I_0 < 600$ mA, the data scales up to $f < 8$ kHz. At $f > 8$ kHz, as seen in Fig. 4(d), the $V'(f)$ deviates from the scaled curve with $\frac{V'}{I_0}$ declining and reaching minima before resuming its increasing trend at higher $f$. The departure from scaling increases as $I_0$ increases. The above suggests a significant non-linear response with varying amplitude of the ac excitation current in the (ii) regime, viz., $T_g < T < T^*$. In this regime, the $V'(f)$ drops rather than increase with $f$, in the range of 8 to 12 kHz, as seen in Fig. 4(d). This non-linear response cannot be attributed to the heating of the sample because then it would have been present in all the three $T$ regimes. Furthermore, our non-contact measurement technique restricts any kind of contact heating and thus ensures the non-linear response coming inherently from the sample itself.

## DISCUSSION

Our measurements in $SmB_6$ enable us to distinguish between three distinct $T$ regimes, viz. (i) $T \geq T^*$ (~ 66 K), (ii) (40 K~) $T_g \leq T < T^*$, and (iii) $T < T_g$. Our $\chi(T)$ measurements in regime (i) show presence of Curie-Weiss behavior with no magnetic ordering. In the $T$ regime (ii) i.e. below $T^*$, the strong correlation effects start to develop and they play an important role in shaping the properties of SmB$_6$. Near $T^*$, both DC magnetization (see Fig. 2(b) inset) and our AC two coil pickup measurements (see Fig. 1(b)) show noticeable change in curvature, and a weak bump like feature in $\frac{d\rho}{dT}$ is found in electrical transport measurements (see supplementary section III). The two-coil pickup signal noticeably drops in regime (ii) and reaches its minimum value at $T_g$. In regime (iii), despite the large increase in the bulk resistivity below $T_g$ our two-coil pickup signal recovers, revealing that the material possesses a thin surface layer of sub-micron thickness with high conductivity. It is worth noting that below $T_g$ the bulk conductivity of the sample diminishes significantly. We observe a distinctive sublinear frequency dependence of the pickup signal in this $T$ regime (iii), i.e., $V'(f) \propto f^{\frac{1}{2}}$. Here we also find a dominant linear $T$ dependence of the pickup signal suggesting strong correlation governing the behavior of surface electrical conductivity below $T_g$. We find that the pick-up response scales with the drive (excitation) amplitude at $T < T_g$ and also at $T > T^*$ (see Fig. 4(c) and Fig. 4(e)) whereas in between $T_g$ and $T^*$ there is a non-scaled regime (see Fig. 4(d)). To understand the scattering mechanism governing electrical transport in regime (i) ($T \geq T^*$), we approximate the behavior of the $\rho(T)$ using a conventional Hamann function (see Fig. 2(a) main panel and inset (I)) of the form, $\rho(T) = \rho_{K0}\left[1 - \frac{\ln(T/T_K)}{\sqrt{\ln^2(T/T_K) + s(s+1)\pi^2}}\right] + \rho_0$, where $T_K$ is the Kondo temperature scale, $s$ is the spin of the magnetic impurity, $\rho_{K0}$ is the proportionality constant and $\rho_0$ is the residual resisitivity [51]. The data in Fig. 2(a) inset (I) fits to this function from high $T$ to $T^*$ with fitting paramters $T_K = 7.80 \pm 0.91$ K, $s = 2.4 \pm 0.1$, $\rho_{K0} = 0.00801 \pm 0.00009$ $\Omega$.cm and $\rho_0 = 0.00229 \pm 0.00004$ $\Omega$.cm. The low $T_K$ value suggests weak exchange interaction between conduction electrons with a dilute density of uncorrelated magnetized impurities. Below $T^*$ the deviation of the measured $\rho(T)$ from the Hamann function fit, shows a failure of single-impurity model in regime (ii). In regime (ii) at $T < T^*$ there is the gradual onset of Kondo hybridization between the Sm ion moments and the itinerant electrons in SmB$_6$ and one cannot consider anymore the magnetic ions as "dilute", i.e. independent. Below $T^*$ at $T_g = 40$ K, a uniform Kondo gap develops across the bulk of the sample. We recall here that earlier transport measurements in SmB$_6$ [4,5,14,15,49] also reported seeing a bulk Kondo gap opening temperature in the range of 30 to 50 K, which is consistent with our $T_g$ value of 40 K. It's interesting to note here that contrary to us, these earlier transport studies suggest the presence of a high conducting state only at very low $T$, typically well below 10 K. In the context of our regime (ii), Scanning Tunneling Spectroscopy (STS) measurements in SmB$_6$ [50], reported seeing suppression in the density of states and a gap like feature emerging in their spectra from ~ 60 K, which is similar to our $T^*$ scale. The STS spectra shows an evolution with lowering of $T$. The gap like feature which begins developing from 60 K becomes a robust Kondo hybridization gap at 40 K, which is similar to our $T_g$. Furthermore similar to our study, from Point contact spectroscopy one can also identify three different temperature regimes [23]. At $T > 90$ K, the highly symmetric conductance spectra with negligible $T$ dependence is marked as the regime of weak interaction between the electrons and the local moments [23]. The temperature range of 90 K to 30 K is described as a single-ion resonance regime where the zero bias conductance is suppressed and an asymmetry starts to develop in the conductance spectra. Below 30 K, the highly asymmetric conductance spectra suggests the emergence of Kondo hybridization due to strong correlation [23]. Here our regime (i) is consistent with the weak interaction regime reported in the spectroscopy studies [23]. In the regime (ii) i.e. $T_g \leq T < T^*$ the strong $f - d$ exchange interaction at Sm ion sites dominates over the weaker electron-impurity ion exchange interaction (which is present above $T^*$). We would like to mention that there maybe a mismatch between our temperature scales and those reported in literature [1,5,15,23,30,50], which is due to differences in sample quality that affects the location of the temperature scales. With the bulk fully gapped at $T_g$, we see the presence of a thin high conducting surface layer contribution to conductivity in the vicinity of $T_g$ in our pick-up coil measurements. It is interesting to note here that Angle resolved photoemission spectroscopy (ARPES) in $SmB_6$ [20] reports observation of dispersive states which is of surface origin within the Kondo hybridization gap. A well-formed Kondo gap feature is seen at 40 K which is similar to our $T_g$ value. The ARPES study [20] further suggests that these states maybe present even at $T$ above 40 K. It also indicates that these states with surface origin possess chirality of the orbital angular momentum [20]. Our studies imply that the thin surface layer we observe at $T < T_g$, possess high surface conductivity. Strong correlation effects modify the scattering, leading to our observed linear $T$ dependence of the surface conductivity below $T_g$. In our $T$ regime (ii), which is between $T^*$ and $T_g$, it is likely that the interplay of the emerging bulk Kondo hybridized state and the emerging highly conductive strongly correlated surface layer conductivity, results in the observed non-scaling behavior of $V'(f)/I_0$ (cf. Fig. 4(d)). Our findings and the discussion above lead us to consider that $SmB_6$ has signatures of weak interactions above $T^*$, where the transport is dominated by scattering from dilute magnetic ions. At $T < T^*$ a strongly correlated Kondo hybridization state between the localized Sm ion moments and the itinerant electrons, progressively takes hold. At $T_g$, a Kondo hybridization gap uniformly opens throughout the majority of the sample. It is probable that the strongly correlated surface layer response is present above $T_g$ and it begins to affect the conductivity of $SmB_6$ between $T^*$ and $T_g$. However, we clearly detect the emergence of a thin high conducting surface layer in $SmB_6$ from just below $T_g$ when the bulk electrical conductivity rapidly diminishes due to the opening of the Kondo gap in bulk of the material.

## CONCLUSION

In conclusion, through careful comparison with a conventional TI material $Bi_2Se_3$, we identify three distinct temperature regimes of bulk Kondo gap formation in $SmB_6$. The process sets in from below a temperature scale $T^*$ and the uniform gap appears across the entire bulk only below $T_g$. Simultaneously, in the vicinity of $T_g$, the features of a strongly correlated thin high conducting surface layer begins to emerge in $SmB_6$. More detailed experimental and theoretical investigations are needed in the future to probe these regimes in strongly interacting TI materials.

**Acknowledgements:** The authors thank Priscila. F. S. Rosa (Los Alamos National Laboratory, Los Alamos, New Mexico 87545, USA) and A. Bharathi (UGC-DAE Consortium for Scientific Research, Kalpakkam 603104, India) for single crystals. We also thank P.F.S. Rosa and S.M. Thomas for private communication and fruitful discussions. SSB thanks funding support from DST-SERB (SUPRA) and IIT Kanpur. SG thanks CSIR, INDIA for funding support. SP thanks the Prime minister's Research Fellowship (PMRF) Scheme of the Ministry of Human Resource Development, Govt. of India, for funding support.

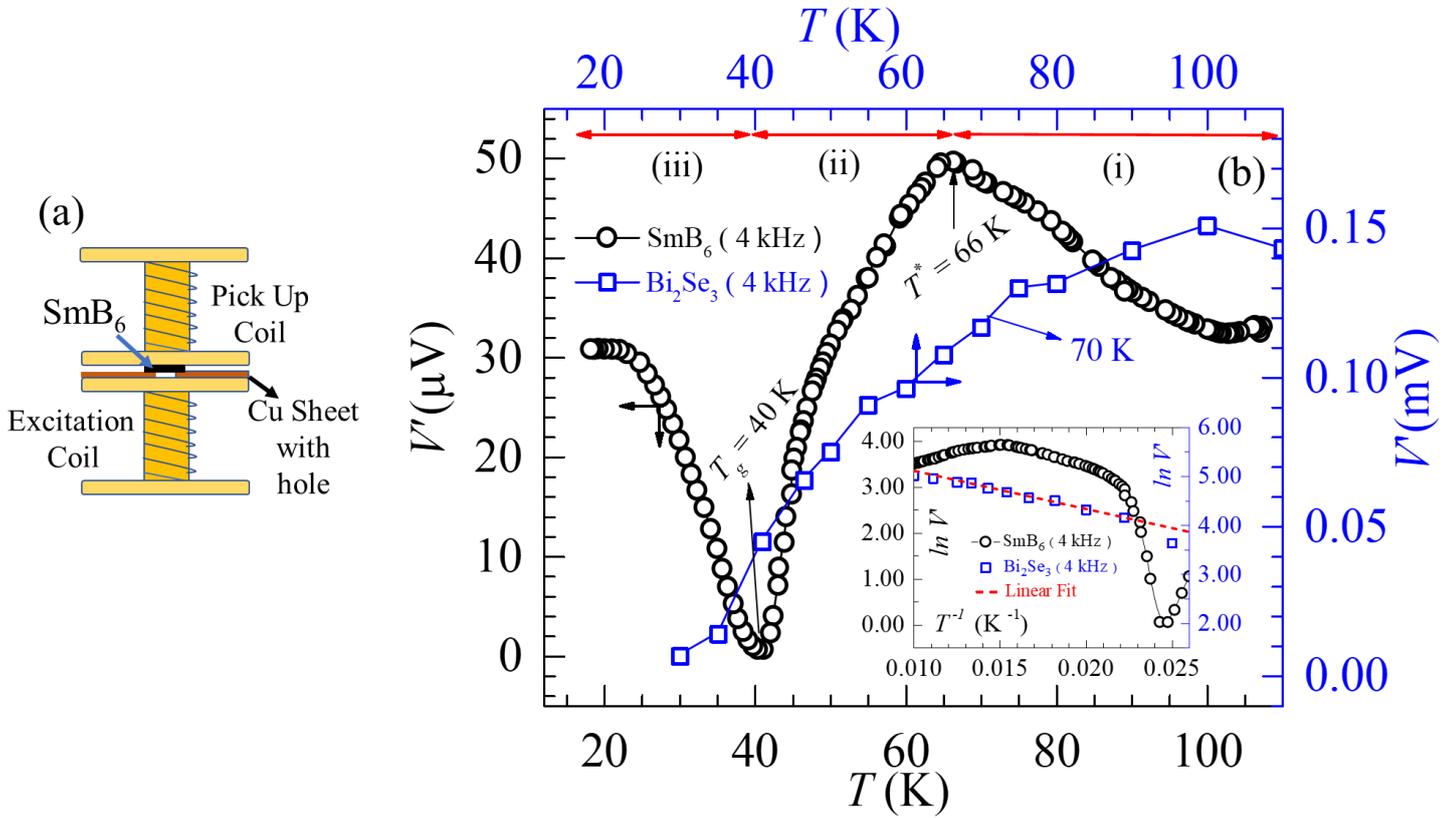

**Fig.1. Schematic of the setup and temperature dependence of the in-phase pick-up signal ($V'$) of SmB$_6$ and Bi$_2$Se$_3$.** (a) The schematic of our two-coil setup with SmB$_6$ placed between the coils. (b) The $V'$ vs $T$ data for SmB$_6$ (black open circles) and Bi$_2$Se$_3$ (blue open squares) both for 4 kHz excitation frequency and 200 mA excitation current ($I_0$). Note that the left (black colored) and bottom (black colored) axis correspond to the $V'(T)$ data for SmB$_6$ only and the right (blue colored) and top (blue colored) axis correspond to that for Bi$_2$Se$_3$. The three distinct temperature regimes (i), (ii) and (iii) for SmB$_6$ are marked by red arrows. Inset shows the $ln\ V'$ vs $T^{-1}$ plot for both SmB$_6$ (black open circles) and Bi$_2$Se$_3$ (blue open squares). Here the left and right vertical axis correspond to SmB$_6$ and Bi$_2$Se$_3$ respectively (with a common x-axis). The linear fit (red dashed line) shows the activated nature of the conductivity of Bi$_2$Se$_3$ whereas a completely different nature is observed for SmB$_6$.

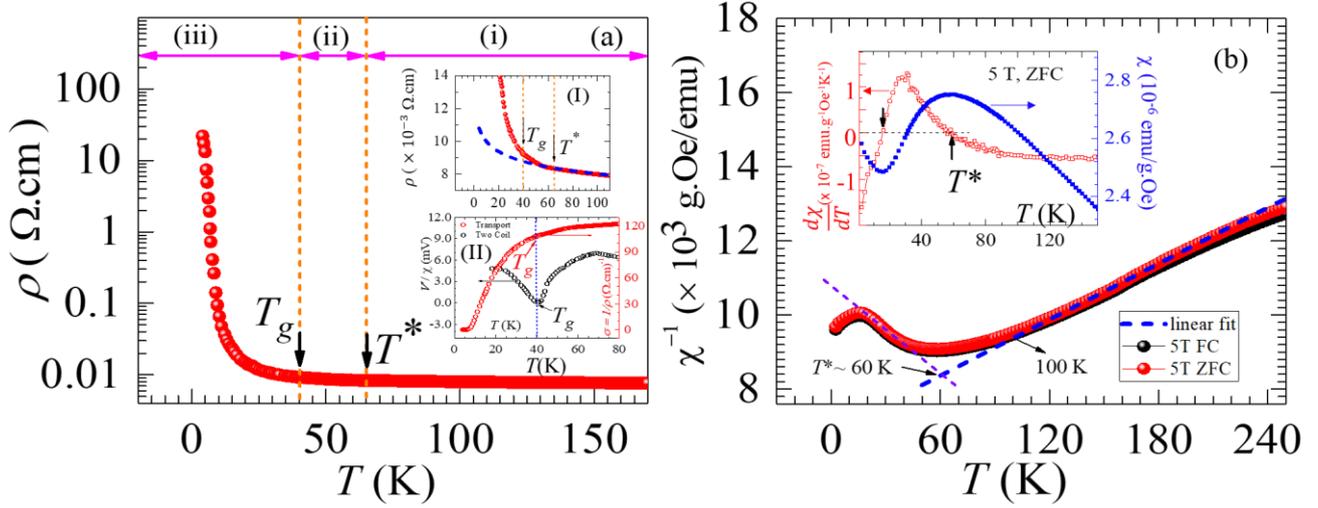

**Figure 2. Transport and DC susceptibility of SmB$_6$ single crystal** (a) Four-probe based electrical resistivity ($\rho$) versus $T$ data for SmB$_6$ single crystal. The figure identifies the three regimes (i), (ii) and (iii) marked by magenta arrows and two temperatures $T_g$ and $T^*$. Inset (I) shows the behavior (blue dashed line) of the Hamann function $\rho(T) \propto \left[1 - \frac{\ln(T/T_K)}{\sqrt{\ln^2(T/T_K) + s(s+1)\pi^2}}\right]$, which fits to the high $T$ regime ($T > T^*$) of the data. One sees the clear departure of $\rho(T)$ from Hamann fit from below $\sim T^*$ (see discussion section of main text for additional details). Inset (II) shows $V'(T)/\chi(T)$ response at 4 kHz excitation frequency (black circles) and bulk DC conductivity ($\sigma$) response (red circles) of SmB$_6$ from two coil and transport experiments respectively. The $\sigma$ starts to decrease at $T_g$ = 40 K whereas $V'(T)/\chi(T)$ shows significant increase from $T_g$ = 40 K.(b) DC magnetic susceptibility measurements on SmB$_6$ shows $\chi^{-1}$ vs $T$ plot for both ZFC (red spheres) and FC (black spheres) under 5 T magnetic field. Blue dashed line shows the Curie Weiss fit above 100 K. Linear back extrapolation (violet dashed line) marks the $T^* \sim$ 60 K. Inset shows $\frac{d\chi}{dT}$ vs $T$ (red open squares and left y axis) and $\chi$ vs $T$ (blue closed squares and right y axis) for 5 T ZFC. $T^*$ is denoted by black arrows in both the main panel and the inset.

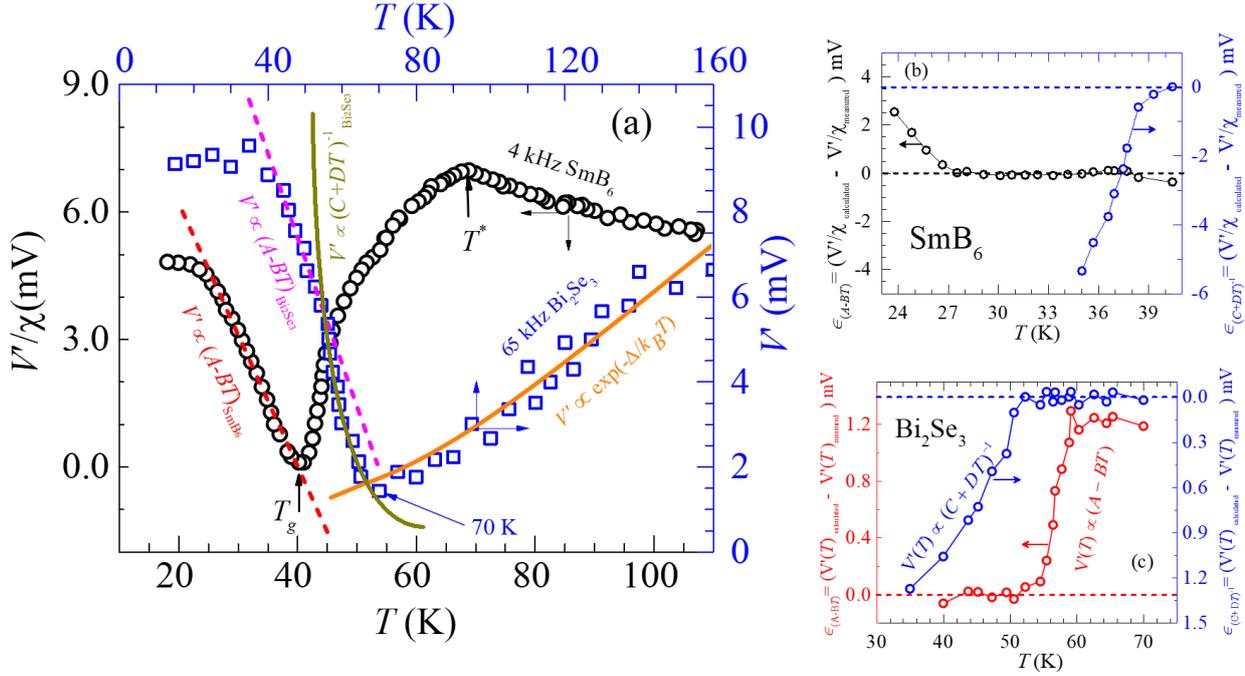

**Fig.3. Comparison of $V'(T)$ of Bi$_2$Se$_3$ and SmB$_6$.** (a) $V'(T)$ response of Bi$_2$Se$_3$ and $V'(T)/\chi(T)$ of SmB$_6$ are denoted by blue open squares and black open circles respectively. Note that the left (black colored) and bottom (black colored) axis correspond to the $V'(T)/\chi(T)$ data for SmB$_6$ only and the right (blue colored) and top (blue colored) axis correspond to $V'(T)$ data for Bi$_2$Se$_3$. Dark yellow fit shows the $V'(T) \propto 1/(C + DT)$ behavior of Bi$_2$Se$_3$ surface state conductivity. Orange solid line fit shows $V'(T) \propto \exp\left(-\frac{\Delta}{k_BT}\right)$ behavior of Bi$_2$Se$_3$ bulk state conductivity. Red and magenta dashed lines show distinct $V'(T)/\chi(T) \propto (A - BT)$ and $V'(T) \propto (A - BT)$ behavior of SmB$_6$ and Bi$_2$Se$_3$ surface conductivity respectively. (b) show the difference between the $V'(T)/\chi(T)$ data calculated from the fits ($\frac{1}{C+DT}$ and $(A - BT)$) and from experiment for SmB$_6$ and (c) the difference between the $V'(T)$ data calculated from the fits ($\frac{1}{C+DT}$ and $(A - BT)$) and from experiment for Bi$_2$Se$_3$.

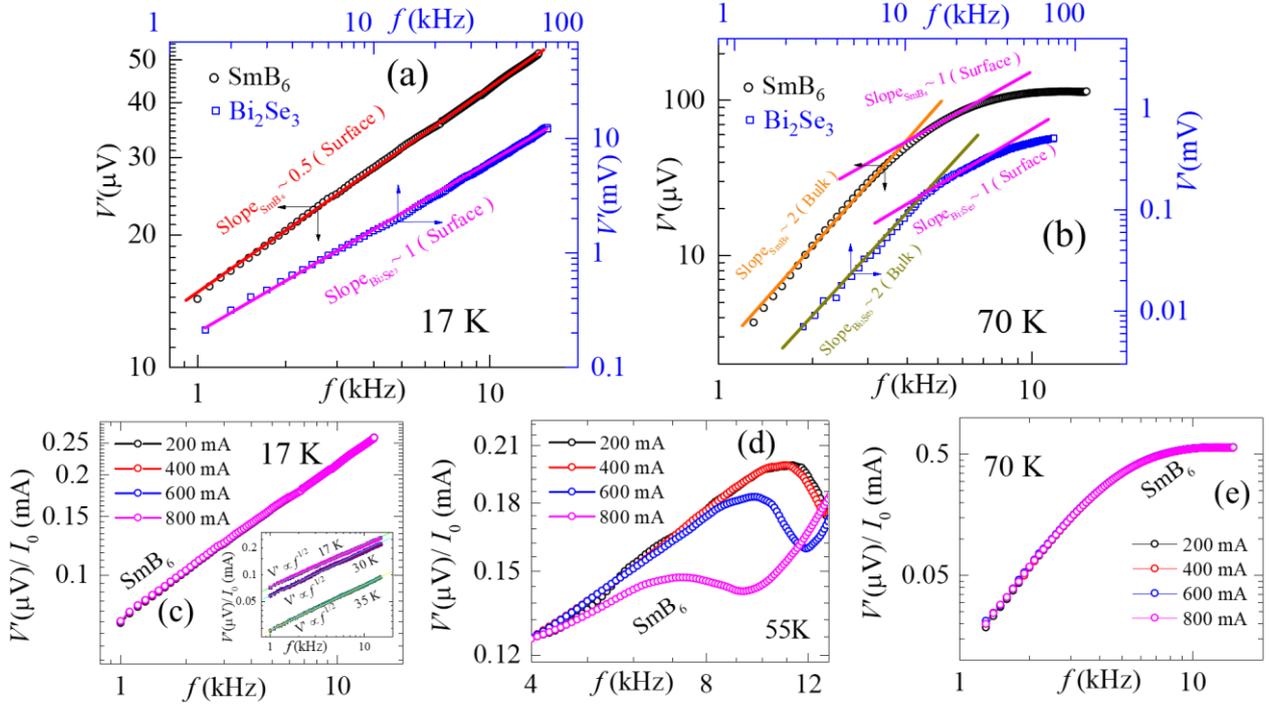

**Fig.4. Excitation frequency and $I_0$ dependence of the pick-up signal of SmB$_6$ and Bi$_2$Se$_3$ at different $T$.** (a) shows the log-log plot of $V'(f)$ response of Bi$_2$Se$_3$ (blue open squares) and SmB$_6$ (black open circles) at 17 K. Red and magenta solid lines indicate the $V' \propto f^{\frac{1}{2}}$ and $V' \propto f$ response of the sample SmB$_6$ and Bi$_2$Se$_3$ respectively. (b) shows the log-log plot of $V'(f)$ response of Bi$_2$Se$_3$ (blue open squares) and SmB$_6$ (black open circles) at 70 K. Orange and dark yellow solid lines indicate the $V' \propto f^2$ response of both the samples SmB$_6$ and Bi$_2$Se$_3$ respectively. Magenta solid lines indicate the $V' \propto f$ response of both of the samples SmB$_6$ and Bi$_2$Se$_3$ (c) shows the scaled behavior of $V'(f)$ response of SmB$_6$ at 17 K for different excitation currents ($I_0$). Inset shows, the scaled behavior of $V' \propto f^{\frac{1}{2}}$ seen at different $T$ of 17 K, 30 K and 35 K, below $T_g$= 40 K. Note that at each $T$, $V'(f)$ is measured with different excitation amplitude $I_0$ and then the normalised $V'/I_0$ data is plotted versus $f$. (d) shows the non-scaled nature of the $V'(f)$ response of SmB$_6$ at 55 K for different excitation currents. (e) shows the scaled behavior of $V'(f)$ response of SmB$_6$ at 70 K for different excitation currents. Note that the excitation current values shown in the figures correspond to the peak value ($I_0$) of the AC current in the primary excitation coil.

# Supplementary Information

## Exploration of strongly correlated states in SmB$_6$ through a comparison of its two-coil pick-up response to that of Bi$_2$Se$_3$

Sayantan Ghosh[1], Sugata Paul[1], Amit Jash[1,2], Zachary Fisk[3], S. S. Banerjee[1,†]

[1]*Indian Institute of Technology Kanpur, Kanpur, Uttar Pradesh 208016, India.*
[2]*Department of Condensed Matter Physics, Weizmann Institute of Science, Rehovot, Israel.*
[3]*Department of Physics and Astronomy, University of California at Irvine, Irvine, CA 92697, USA.*

Corresponding author email : [†]satyajit@iitk.ac.in

**SECTION I:**

**Details of two-coil setup**

In our two-coil setup, both the excitation coil and pick-up coil are dipole coils [1,2]. The winding of the excitation and pick-up coils should be as close as possible. The excitation coil has four layers and each layer has 36 turns. The pick-up coil has four layers having 32 turns for each layer. Due to these closely matched parameters of the coils, we can extract the response of the sample by simply subtracting the background response from that of sample with background. Fig. 1 shows the image of our setup. The bobbin and stand are made of an insulating and non-magnetic material Macor to avoid any unwanted signals developed by eddy currents. The sample is kept in between the two coils.

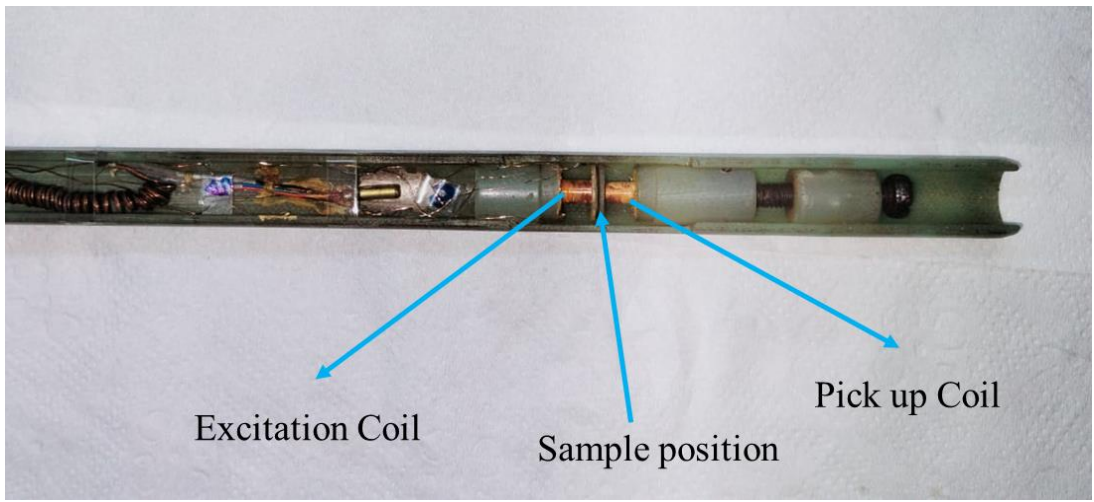

**Fig.1. Two-coil experimental setup**

**SECTION II:**

**XRD measurement of SmB$_6$ single crystal**

Fig. 2 shows the XRD pattern of our SmB$_6$ single crystal. We obtained four peaks i.e. (100), (200), (300) and (400) at $2\theta$ positions around 21.49°, 43.9°, 68.02° and 96.61° respectively. These peaks were verified by comparing with the PCD file data for standard powder XRD pattern of SmB$_6$ (refer to Pearson's crystal database (PCD) file no. 1126045). Absence of any additional impurity phases in the XRD pattern ensured that the sample is single phased and a single crystal with growth direction along (100) crystal plane. The Bragg's diffraction condition is given by:

$$2d\sin\theta = n\lambda \text{ ; where } d = a/\sqrt{h^2 + k^2 + l^2}.$$

Here $d$ is the separation between the crystal planes with $a$ being the lattice constant for cubic lattice, $(h\ k\ l)$ are the Miller indices of the plane and $n$ is the order of diffraction. Using $\lambda = 0.15406$ nm as the standard X-ray wavelength (Cu, $K_{\alpha 1}$ line) for the (100) peak, we obtain $a = 0.413$ nm. This value exactly matches with the previously published data for SmB$_6$ [3,4].

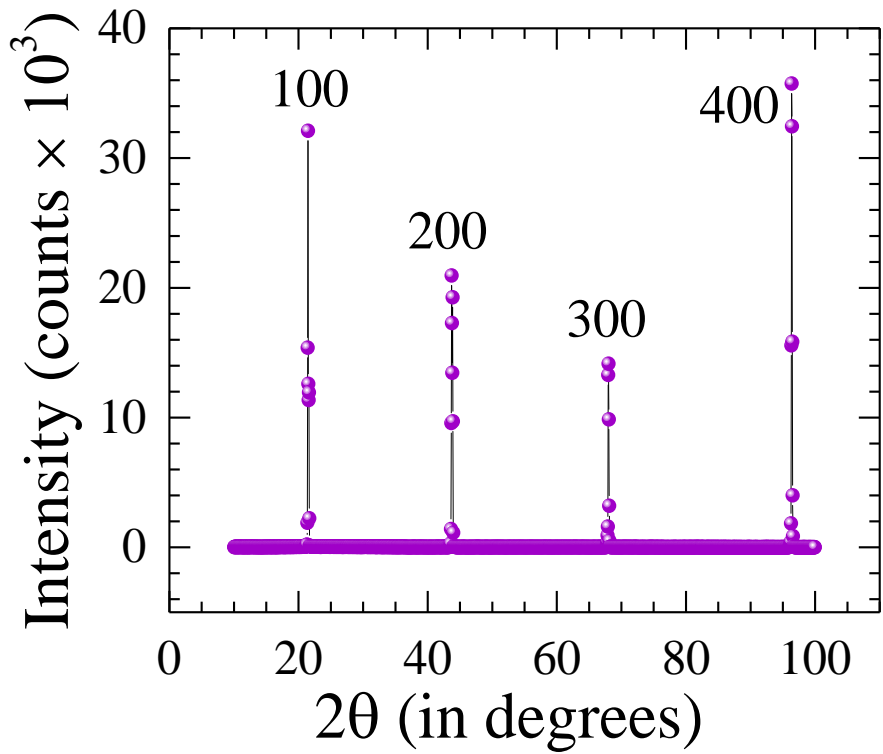

**Fig. 2. XRD of SmB$_6$ single crystal**

## SECTION III:

## $\frac{d\rho}{dT}$ vs $T$ data for SmB$_6$

Fig. 3 shows the $|\frac{d\rho}{dT}|$ vs $T$ plot for SmB$_6$ obtained from the four-probe transport measurements. We can observe a change in the nature of the plot at two different temperatures (marked by black arrows in Fig. 3) i.e. near $T^* = 65$ K $\pm$ 2 K and $T_g = 40$ K $\pm$ 2 K.

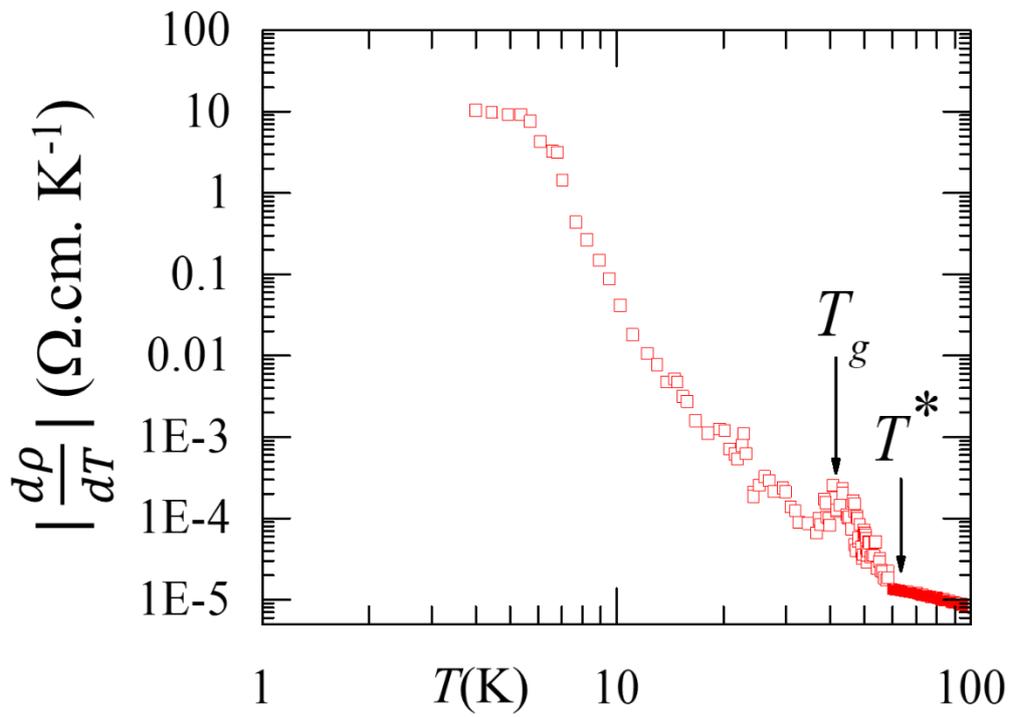

**Fig.3.** $|\frac{d\rho}{dT}|$ vs $T$ plot for SmB$_6$ in log-log scale

## SECTION IV:

### *M - T* response of SmB$_6$

Fig. 4 shows the DC magnetization response of SmB$_6$ for both zero field cooled (ZFC) and field cooled (FC) conditions at 5 T magnetic field. No bifurcation between the ZFC and FC data indicates absence of irreversible or history dependent magnetism in SmB$_6$. The dome shaped feature around 60 K suggests the onset of hybridization by Kondo screening of magnetic moments leading to a strongly correlated gap formation in the bulk (see main text for detailed explanation). The upturn below 15 K may be attributed to magnetic in-gap states present inside the sample. The data is excellently consistent with previously reported magnetization data by Biswas *et al* [5].

Note that the $\chi(T)$ used in all the $V'(T)/\chi(T)$ plots of main manuscript are dimensionless (values of dimensionless $\chi(T)$ are calculated from the $\chi(T)$ (in emu/g.Oe units) plot below).

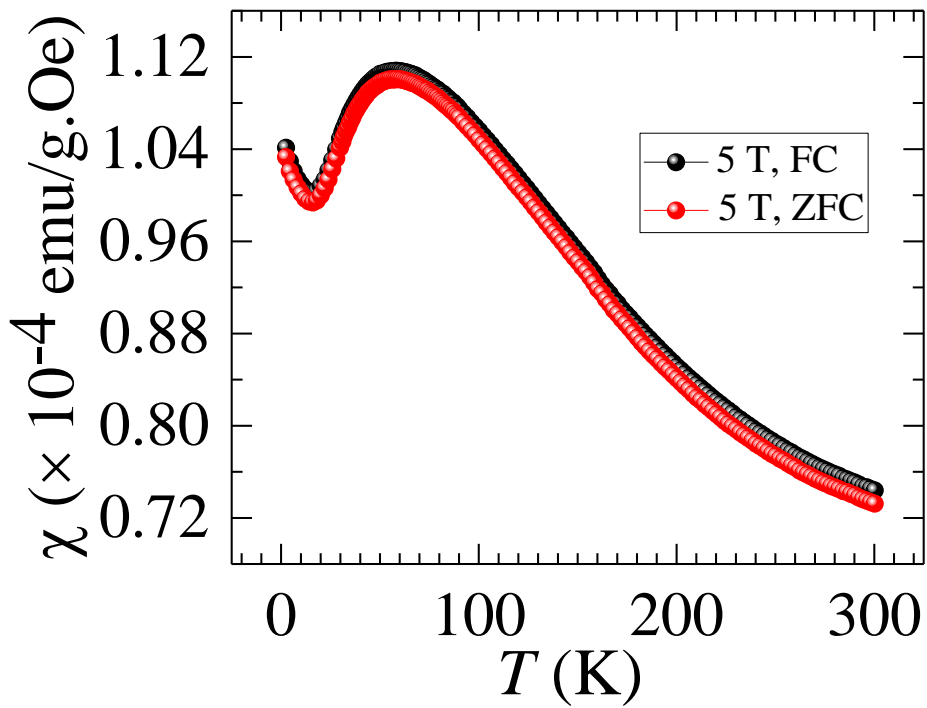

**Fig.4. DC magnetic susceptibility $\chi(T)$ response of SmB$_6$**

## SECTION V:

## M - H loop of SmB₆

Figure below shows the $M$ - $H$ loop of $SmB_6$ at different temperatures for the entire field range with values up to 6 T. No feature of non-linearity and Hysteresis was observed at any temperature, showing no history dependent magnetic order is present inside the sample. This is similar to the $M$ - $H$ behavior of $SmB_6$ reported earlier by P. K. Biswas *et al*. [5]. Inset of Fig 5 also shows the low field $M$ - $H$ response of $SmB_6$ up to 100 mT field at different temperatures.

The $M(H)$ is linear at both low and high field ($H$) regimes but with slightly different slopes, of $1.5\times10^{-4}$ emu/(g.Oe) and $1\times10^{-4}$ emu/(g.Oe), respectively. Therefore, the difference between the susceptibilities at high field (~ 5 T) and in the typical low magnetic field of the two-coil measurement (~ 20 mT) is in the range of $5\times10^{-5}$ emu/(g.Oe), which is not very significant. This suggests, there cannot be two distinct magnetic phases in these two field regimes.

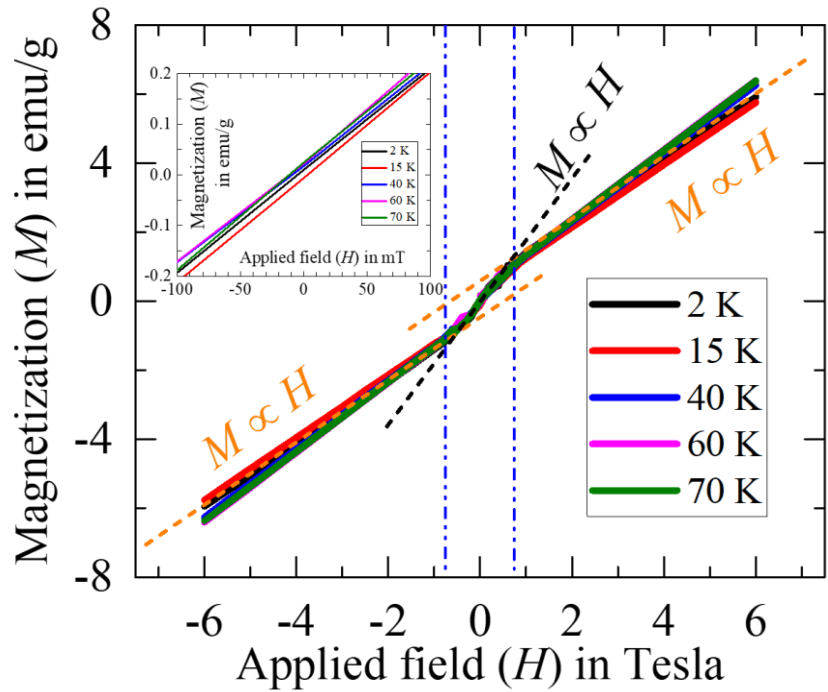

**Fig. 5.** $M$ - $H$ loop of $SmB_6$ at different $T$ upto 6 T field (inset shows the low field $M$ – $H$ behavior up to 100 mT field)

## SECTION VI:

**Comparison of $\frac{V'}{\chi}$ and $\sigma$ vs $T$ up to 300 K in SmB$_6$**

Figure 6 below shows the comparison of $V'(T)/\chi(T)$ data ($V'(T)$ response is at 4 kHz excitation frequency) with the bulk conductivity $\sigma(T)$ data (from transport) in SmB$_6$ up to 300 K. Pick-up response $V'(T)$ is proportional to both electrical conductivity ($\sigma(T)$) and magnetic susceptibility ($\chi(T)$). Hence, the normalised $V'(T)/\chi(T)$ is now mainly dependent on $\sigma(T)$ and can be directly compared to the electrical conductivity data. Note that the size of the cusp like feature at $T^*$ is significantly diminished compared to the rise in $V'(T)/\chi(T)$ seen below $T_g$ (see Fig 6 below). Therefore, this difference suggests the enhancement in conductivity of the thin surface layer below $T_g$. At high $T$ up to 290 K, as the bulk signal fully takes over the pick-up response, $V'(T)/\chi(T)$ and $\sigma(T)$ both show a similar nearly-saturating feature. Note that apart from the small change in $\chi(T)$ in SmB$_6$, at fixed $T$, the $\chi$ also doesn't show any significant frequency dependence. Therefore, normalization with $\chi$ doesn't change the feature of the pickup signal $V'(f)$ we have reported in our paper.

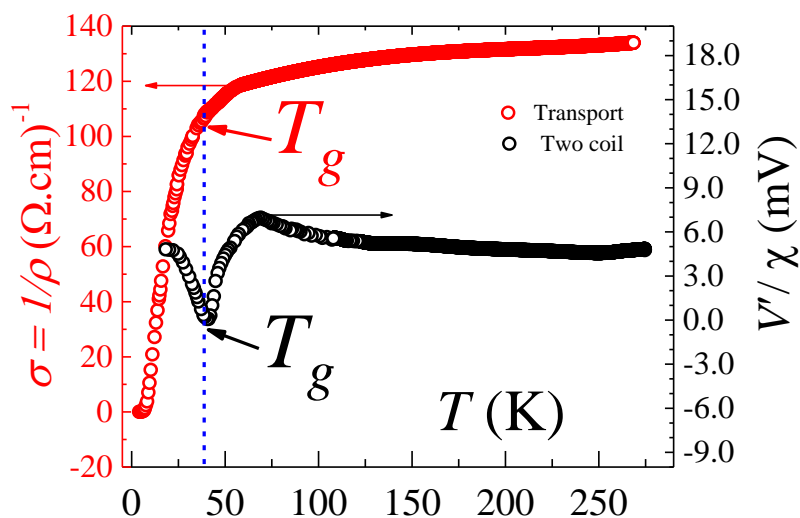

**Fig. 6.** $\frac{V'}{\chi}$ and $\sigma$ vs $T$ in SmB$_6$ up to 290 K

## SECTION VII:

## Two-coil measurements of SmB$_6$ at 6 kHz excitation frequency

Fig. 7 shows the $V'$ vs $T$ data for 6 kHz excitation frequency and 200 mA excitation current in SmB$_6$. The 6 kHz response is almost identical to the 4 kHz response (see main text). $V'$ shows a peak at around $T^* \sim 66$ K and then it starts to decrease to finally show a minima at $T_g \sim 40$ K. Below $T_g$ the $V'(T)$ again shows an upturn indicating the emergence of a thin high-conducting surface layer (see main text for details).

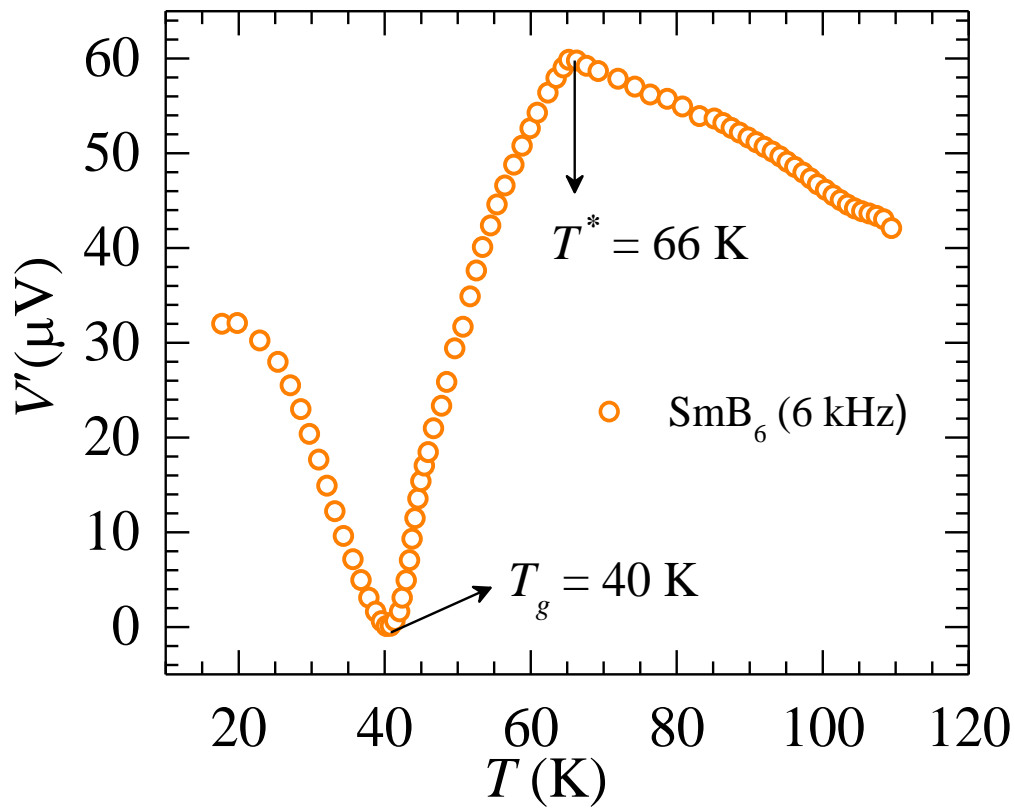

**Fig.7. $V'$ vs $T$ data of SmB$_6$ for 6 kHz excitation frequency and 200 mA excitation current**

## SECTION VIII:

**Surface conductivity of SmB$_6$ at 4, 8, 12 and 55 kHz excitation frequency**

Fig. 8 shows the $V'(T)$ response of SmB$_6$ at a varied range of excitation frequencies i.e. 4, 8, 12 and 55 kHz with 200 mA excitation current. Here the $V'(T)$ data is normalised with respect to the response at the lowest temperature i.e. $V'(17\ \text{K})$. Note that the upturn of $V'(T)$ signifying thin surface layer conductivity below $T_g \sim 40$ K is present for both low frequency (4 – 12 kHz) and high frequency (55 kHz) regimes. Hence it is observed that the response from the surface conductivity of SmB$_6$ is achieved even by the low excitation frequency i.e. 4 kHz whereas for Bi$_2$Se$_3$ the upturn of $V'(T)$ was observed at only high frequency excitation, i.e. 65 kHz (see Fig. 3(a) of main manuscript).

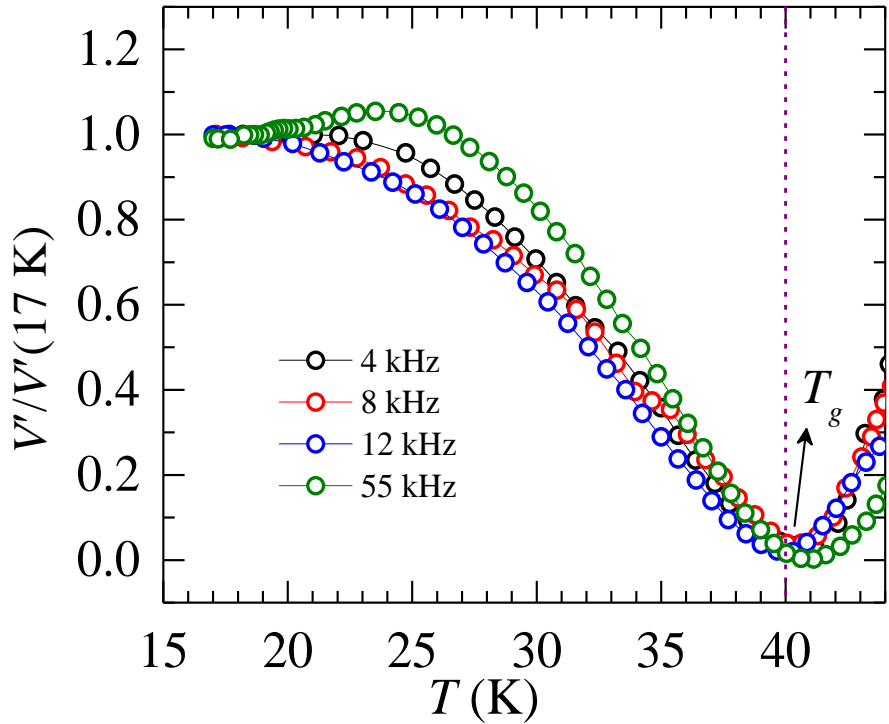

**Fig. 8.** $V'$ vs $T$ data of SmB$_6$ for 4, 8, 12 and 55 kHz excitation frequency and 200 mA excitation current. The $V'(T)$ data is normalised by $V'(17\ \text{K})$ for each frequency.

## SECTION IX:

$V'(f) \propto f^{\frac{1}{2}}$ response of SmB$_6$ below $T_g$

Figure 9 below shows the $V'(f) \propto f^{\frac{1}{2}}$ behavior of SmB$_6$ observed at different temperatures (17 K, 27 K, 30 K, 33 K and 35 K) below $T_g$ (= 40 K). At each temperature, the $V'(f)$ signal is obtained at a constant excitation current amplitude $I_0$= 200 mA. This strongly supports that the unique $f^{\frac{1}{2}}$ behavior is present in the entire $T < T_g$ regime of SmB$_6$ where the response of the thin high conducting surface layer gets modified by strong correlation. Note that, the $V'(f)$ response at all these different temperatures below $T_g$ is found to be scaled with different values of values $I_0$ and all of them shows $V'(f) \propto f^{\frac{1}{2}}$ response (as seen in Fig. 4(c) inset of main manuscript).

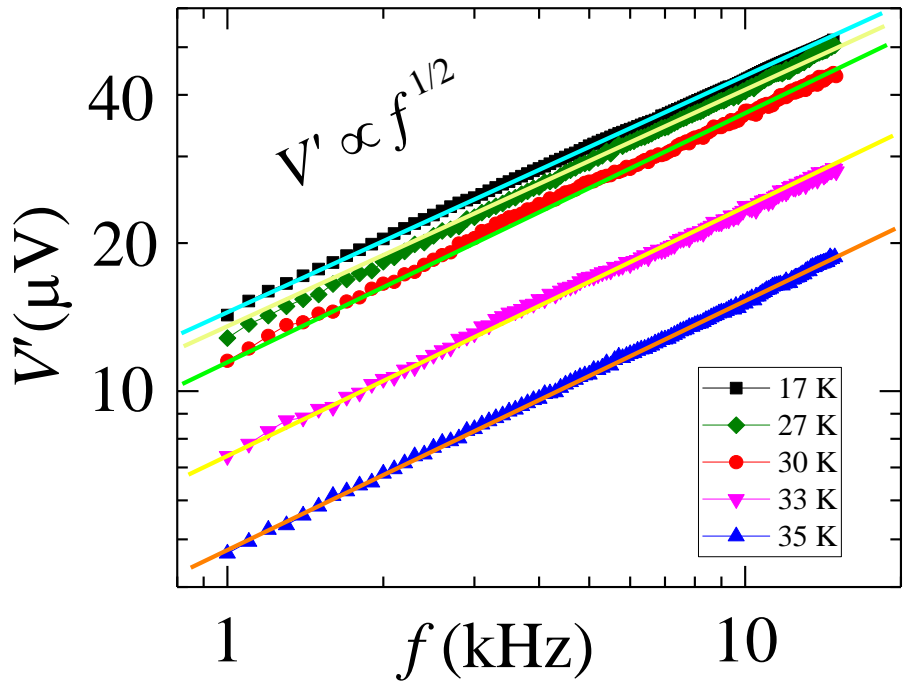

**Fig. 9.** $V'(f) \propto f^{\frac{1}{2}}$ **data of SmB$_6$ at different $T < T_g$ for a constant excitation current amplitude ($I_0$) 200 mA.**

# SECTION X:

## Sample and Background Response for SmB$_6$

Fig. 10 shows the $V'(f)$ response of SmB$_6$ from two-coil measurements separately for 'with sample' (i.e. sample + background combined response) and 'without sample' (only background response) for 200 mA and 400 mA excitation current at 17 K.

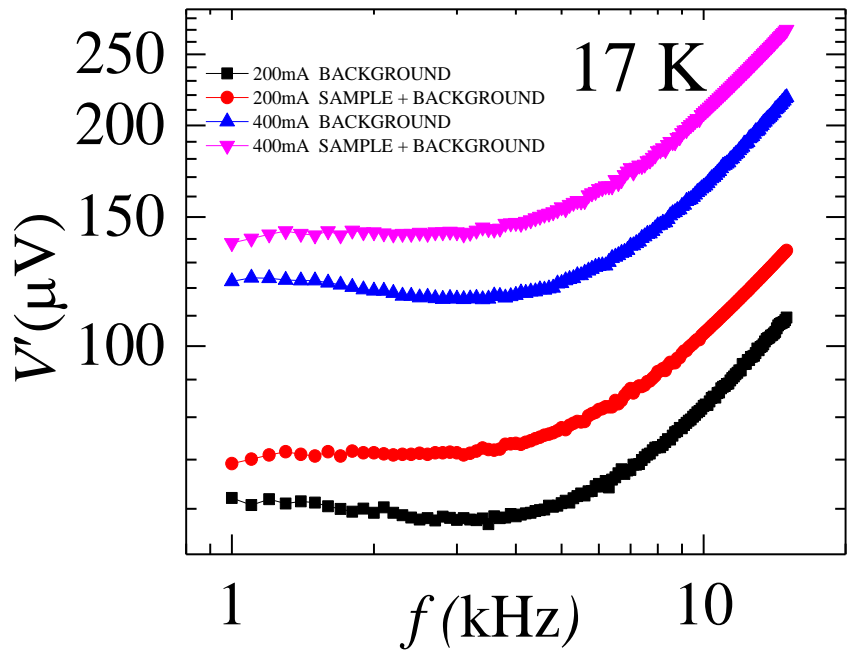

**Fig.10.** $V'(f)$ response of SmB$_6$ 'with Sample' (sample + background) and 'without sample' (background) at 17 K for 200mA and 400mA excitation current

# SECTION XI:

## Characterization of Bi₂Se₃ single crystal

The Bi$_2$Se$_3$ single crystal used for the experiments in our paper is from the same batch of samples reported in [1] and [2]. For transport characterization on this sample, the Fig. 11(a) inset shows the longitudinal resistance $R_{XX}$ of Bi$_2$Se$_3$ single crystal as a function of magnetic field $B$ measured in the standard Van der Pauw geometry at different temperatures of 3 K, 4.2 K, 7 K, 10 K, and 25 K (legends as given in the main panel). In the main panel of Fig. 11(a), $\Delta R_{XX}$ vs $1/B$ plot is shown that indicates SdH oscillations in Bi$_2$Se$_3$ at these different temperatures. A polynomial fit $R_{\text{poly}}(B)$ to the data $R_{XX}(B)$ is taken where $R_{\text{poly}}(B) = R_0 + R_1 B + R_2 B^2$ ($B$: magnetic field, $R_0, R_1, R_2$ are constant fitting parameters). The fit provides $R_0 = 9.26 \times 10^{-3}$ Ω, $R_1 = -1.71 \times 10^{-6}$ Ω·T$^{-1}$, $R_2 = 2.91 \times 10^{-5}$ Ω·T$^{-2}$. $\Delta R_{XX}$ is calculated by subtracting the fitted $R_{\text{poly}}(B)$ values from the experimental dataset $R_{XX}(B)$. The black solid line represents the fit to the LK equation on the $\Delta R_{XX}$ data at 4.2 K. For fitting to the magnetoresistance data in Fig. 11(a), following Ref. [6] and the values therein, we use

$$\Delta R_{XX} = \frac{a}{\sqrt{0.011B}} \left[ \left(\frac{11.12}{B}\right) / \sinh\left(\frac{11.12}{B}\right) \right] e^{-\left(\frac{19.38}{B}\right)} (0.95)\cos\left[2\pi \left\{\left(\frac{F}{B}\right) + \beta\right\}\right].$$

In this equation, $a = 0.0135$ Ω and F and β are the fitting parameters. The solid black line fit gives $F = 46.95 \pm 0.25$ T and $\beta = 0.43$. The Hall measurements on Bi$_2$Se$_3$ in Fig. 11(b) also show the transverse resistance $R_{XY}$ as a function of $B$ at different temperatures of 3 K, 4.2 K, 7 K, 10 K, 25 K (all legends given in Fig. 11(b)). The Bi$_2$Se$_3$ single crystal used in all the above measurements has a thickness of approximately 69 μm.

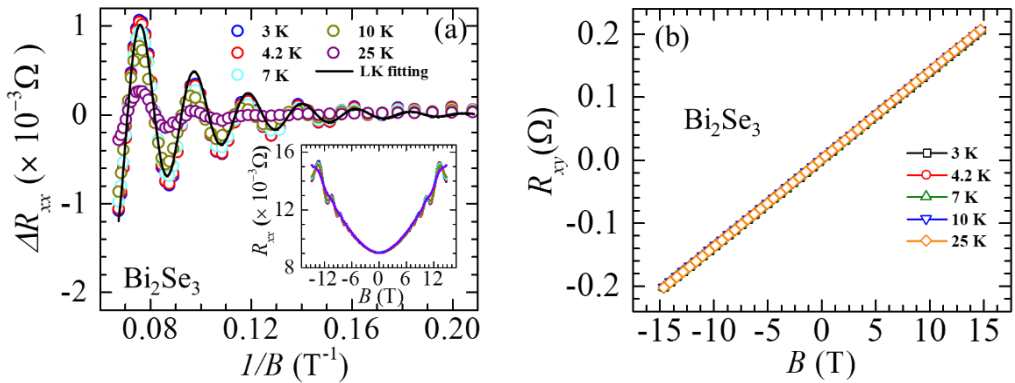

**Fig. 11.** (a) $\Delta R_{XX}$ vs $1/B$ plot of Bi$_2$Se$_3$ single crystal at different temperatures of 3 K, 4.2 K, 7 K, 10 K, and 25 K. Inset shows the $R_{XX}(B)$ data at these temperatures. (b) $R_{XY}(B)$ plot of Bi$_2$Se$_3$ single crystal shown for different temperatures.

## SECTION XII:

## Estimation of skin depth

We have performed a rough estimation of skin depth of the two samples $SmB_6$ and $Bi_2Se_3$ at the two excitation frequencies 4 kHz and 65 kHz respectively. Using the formula of skin depth i.e.,

$$\delta = \sqrt{\rho/\pi f \mu}$$

where $\rho$ = bulk resistivity of the sample, $f$ = AC excitation frequency, $\mu = \mu_0(1 + \chi)$, $\chi$ = susceptibility of the sample (dimensionless), we have estimated the value of $\delta$ at $f = 4$ kHz to be ~ 25 mm at 100 K for $SmB_6$ ($\rho = 10^{-3}$ Ω.cm and $\chi = 6.7 \times 10^{-3}$ values are calculated from bulk transport and susceptibility data respectively). We have chosen the high $T = 100$ K because in this regime the bulk conductivity is dominant. As the calculated $\delta$ ~ 25 mm is much greater than the sample thickness i.e., 0.2 mm, the maximum depth up to which the AC signal can probe at such high $T$ in $SmB_6$ is the thickness of the sample itself. By similar means, we can also assume that even for $Bi_2Se_3$ the maximum depth for the AC signal to penetrate into the sample is its thickness i.e., 0.069 mm for our case at high $T \sim 150$ K.

We know, the value and frequency variation of $\delta$ is mainly governed by high conducting surface states of a topological insulator (TI) material [1]. Hence at $T > 40$ K for $SmB_6$ and $T > 70$ K for $Bi_2Se_3$, where the surface conductivity has not been effective yet, we can roughly consider that the pick-up response $V' \propto$ sample volume. Thus, for a particular sample at higher $T$, $V' \propto$ effective sample thickness up to which the AC signal can impinge into (skin depth $\delta$ crudely). Also, it is known that for $SmB_6$ $V' \propto \chi$. For both $SmB_6$ and $Bi_2Se_3$, we have assumed the validity of the relation $V' \propto \delta.\chi$ from high $T$ down to the $T$ range below which surface conductivity dominates, i.e. 40 K for $SmB_6$ and 70 K for $Bi_2Se_3$. From this assumption we compare the $V'$ values of $SmB_6$ at 4 kHz between 100 K and 40 K from the relation:

$$\frac{V'^{40\,K}_{SmB_6}}{V'^{100\,K}_{SmB_6}} = \frac{\delta^{40\,K}_{SmB_6}}{\delta^{100\,K}_{SmB_6}} \times \frac{\chi^{40\,K}_{SmB_6}}{\chi^{100\,K}_{SmB_6}}$$

and we obtain $\delta_{SmB_6}^{40\,K} \sim 4.8\ \mu m$ ($V'$ data taken from two-coil data and $\delta_{SmB_6}^{100\,K} \sim 200\ \mu m$ taken roughly as the sample thickness). Note in the above relation, $\frac{\chi_{SmB_6}^{40\,K}}{\chi_{SmB_6}^{100\,K}} \sim 1$ as $\chi_{SmB_6}^{40\,K} = 1.076 \times 10^{-4}$ emu/g.Oe, $\chi_{SmB_6}^{100\,K} = 1.05 \times 10^{-4}$ emu/g.Oe.

At $T < 40$ K in SmB$_6$ for 4 kHz, the variation of effective $\delta$ with $T$ is mainly dominated by the conductivity of the thin high-conducting surface layer ($\sigma_s(T)$) as shown in the Fig. 12 below with $\delta = \delta_0 \sqrt{\frac{\sigma_0}{\sigma_s(T)}}$ where $\delta_0 = \delta_{SmB_6}^{40\,K}$ and $\sigma_0 = \sigma_{SmB_6}^{40\,K}$. We have used $\sigma_s(T) = \sigma_{0s}(A - BT)$ where $\sigma_{0s}$ is a constant $T$ independent proportionality factor and $A, B$ values are determined from the linear fit to the effective conductivity response ($V'/\chi$) of SmB$_6$ at 4 kHz in $T < T_g$ regime (See Fig. 3(a) of main manuscript). $\delta$ decreases with $T$ and obtains a value ~ 0.4 μm ~ 400 nm at the lowest $T$ (see Fig. 12 below). We assume that the regime where $V'(T)$ saturates, the $\delta$ value is in the order of 400 nm. This indicates that the AC signal is now probing a thin high conducting surface layer of sub-micron order thickness in SmB$_6$ at $f = 4$ kHz.

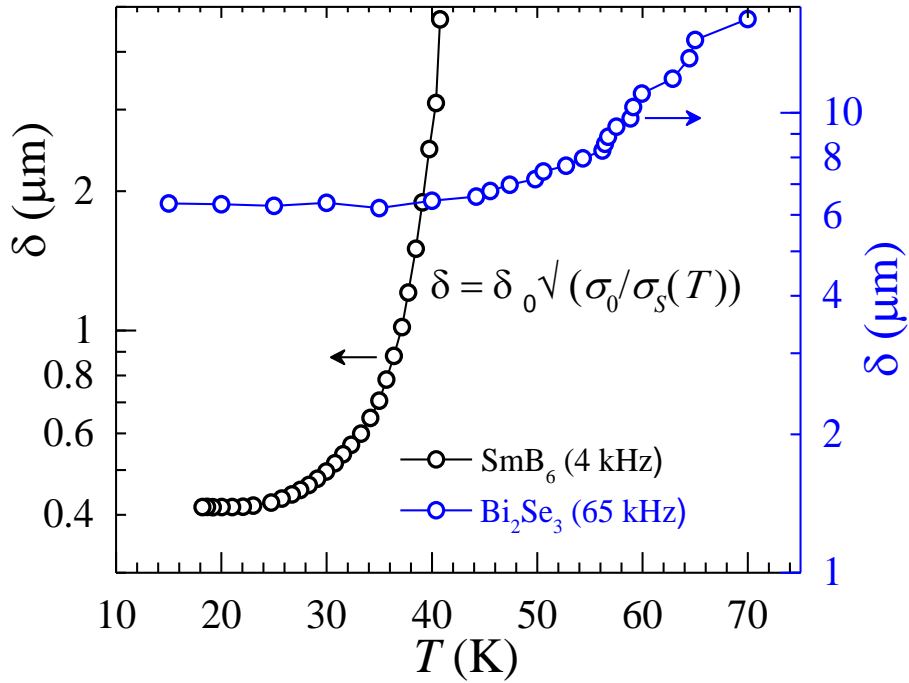

Fig. 12. Skin depth estimation of SmB$_6$ and Bi$_2$Se$_3$

Similarly, for $Bi_2Se_3$ we compare the $V'$ response (taken from two-coil data) at 65 kHz between 150 K and 70 K using the relation

$$\frac{V'^{70\,K}_{Bi_2Se_3}}{V'^{150\,K}_{Bi_2Se_3}} = \frac{\delta^{70\,K}_{Bi_2Se_3}}{\delta^{150\,K}_{Bi_2Se_3}}$$

we obtain $\delta^{70\,K}_{Bi_2Se_3} \sim 16.1\ \mu m$ ($\delta^{150\,K}_{Bi_2Se_3} \sim 69\ \mu m$ taken roughly). Since, $Bi_2Se_3$ is non-magnetic, no $\chi$ dependence arises in this formula. For $T < 70$ K, when high surface state conductivity ($\sigma_s(T)$) starts to dominate hugely in $Bi_2Se_3$, the $\delta(T)$ decreases with $T$ ($\delta = \delta_0 \sqrt{\frac{\sigma_0}{\sigma_s(T)}}$) where $\delta_0 = \delta^{70\,K}_{Bi_2Se_3}$, following the $\sigma_s(T)$ feature for 65 kHz excitation frequency ($\sigma_s(T)$ values are directly obtained from the nature of $V'(T)$ at $T < 70$ K at 65 kHz for $Bi_2Se_3$ in Fig. 3(a) of main manuscript).

## SECTION XIII:

### Surface and bulk fraction of SmB$_6$ and Bi$_2$Se$_3$

Like in Bi$_2$Se$_3$ above 70 K, in SmB$_6$ we know that above 40 K ($T \geq T_g$) there is an admixture of both surface and bulk contributions, and we would like to estimate their relative contributions.

At $T = 70$ K, we estimate the relative fraction of bulk ($P_b$) and surface ($P_s$) contributions to $V'(f)$ by fitting it to $V'(f) = P_b[V_b'(f)] + P_s[V_s'(f)]$ with the constraint $P_b + P_s = 1$. The $V'(f)$ data fits well for SmB$_6$ and Bi$_2$Se$_3$ only in a limited frequency window (see figure 13 below) where contributions from both the surface ($V_s'(f) \propto f$) and bulk ($V_b'(f) \propto f^2$) are present. We therefore, fit the data to a form $V'(f) = mf^2 + nf$ and consider the ratio of the fitting parameters $m:n$ is of the order of $P_b:P_s$. This together with $P_b + P_s = 1$ for the 70 K data, yields for SmB$_6$, $P_b = 70$ % and $P_s = 30$ % and for Bi$_2$Se$_3$, $P_b = 51$ % and $P_s = 49$ %. It is worthwhile noting that for Bi$_2$Se$_3$, the values for $P_b$ and $P_s$ we obtain using the above considerations at 70 K are close to those reported earlier [1].

(Please note that for fitting and comparison we have plotted $V'(f)$ in µV units for both SmB$_6$ and Bi$_2$Se$_3$).

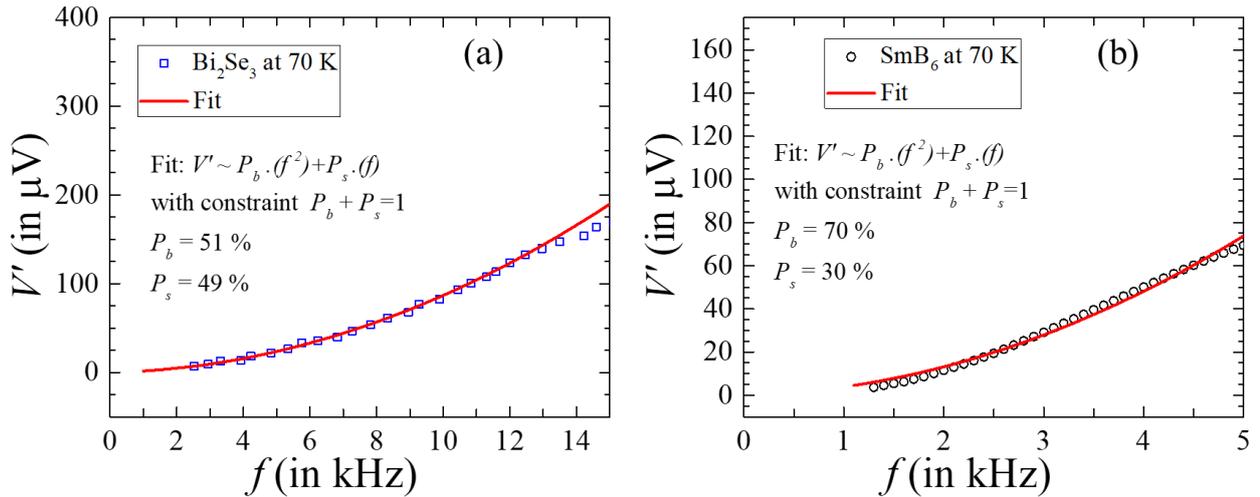

**Fig. 13.** Surface and bulk fraction of (a) Bi$_2$Se$_3$ and (b) SmB$_6$ at 70 K